\documentclass{article}
\usepackage[utf8]{inputenc}
\usepackage{caption}
\usepackage{subcaption}
\usepackage{amsmath}
 \usepackage{cite}
\usepackage{eufrak}
\usepackage{amsfonts}

\newcommand{\sqrts}{\mbox{$\sqrt{\mathrm{s}}$}}

\newcommand{\ppt}{$p_{\rm T}$}

\usepackage{graphicx}

\usepackage{lineno}
\providecommand{\keywords}[1]{\textbf{\textit{Keywords---}} #1}
\usepackage{lineno}

\usepackage{graphicx}
\DeclareGraphicsExtensions{.png,.pdf,.jpeg,.jpg,.eps}
\usepackage{multicol}
\usepackage{float}

\usepackage{lineno}
\begin{document}


\begin{center}
{\Large \bf Application of deep learning in top pair and single top quark production at the LHC}

\vskip1.0cm

Ijaz.Ahmed$^{1}${\footnote{Corresponding author: ijaz.ahmed@riphah.edu.pk}},
Anwar Zada$^{1}$, Muhammad Waqas$^{1}$,
M. U. Ashraf$^{2},$ 
\\

$^1$ Department of Physics, Riphah International University, Islamabad 44000, Pakistan,
\\
$^2$ National Center for Physics, Shahdra Valley Road, Islamabad 44000, Pakistan\\

\end{center}

\vskip1.0cm


\begin{abstract}

We demonstrate the performance of a very efficient tagger applies on hadronically decaying top quark pairs as signal based on deep neural network algorithms and compares with the QCD multi-jet background events. A significant enhancement of performance in boosted top quark events is observed with our limited computing resources. We also compare modern machine learning approaches and perform a multivariate analysis of boosted top-pair as well as single top quark production through weak interaction at    $\sqrt{s}=$14 TeV proton-proton Collider. The most relevant known background processes are incorporated. Through the techniques of Boosted Decision Tree (BDT), likelihood and Multlayer Perceptron (MLP) the  analysis is trained to observe the performance in comparison with the conventional cut based and  count approach.   

\vskip0.5cm
\end{abstract}


\section{Introduction}
The heaviest elementary particle so far discovered is the top quark which is very uncommon quark in many ways. One of the possible fact is, the electroweak decay of top quark is faster than the timescale of hadronisation indicating that it exist only as a free quark. The comparison of precise measurements from the Standard Model (SM) predictions would be helpful in order to understand the possible effects from new physics. More than twenty years ago, the process of ($t\bar t$) pair-production was discovered and this process was entered in the domain of ``precision physics" later after many years~\cite{CDF:1995wbb, D0:1995jca}. Tevatron experiment was the first to observe the production of single top-quark just few years before its shut-down~\cite{D0:2009isq, CDF:2009itk}. Later on the CMS~\cite{CMS:2008xjf} and ATLAS~\cite{ATLAS:2008xda} experiments recorded data at 7 TeV in 2010~\cite{CMS:2011oen} and in early 2011~\cite{ATLAS:2012byx} respectively and ``re-discovered" the production of single top-quark. 

The production of single top-quark at various hadron colliders is useful tool to study and understand the electroweak $Wtb$ vertex and Cabibbo-Kobayashi-Maskawa (CKM) matrix $Vtb$. There are two possible SM production mechanisms are possible at large hadron colliders, the dominant are $t\bar t$ pair production through strong interaction and single top production through electroweak interaction. Single top production, despite being restrained by the weak
coupling constant in comparison to strong $t\bar t$ production has a large cross section
because it is kinematically enhanced. Three different production mechanisms can be studied, first one is known as the t-channel, a virtual t-channel $W$ boson mediates the production of the top quark along with another quark $q b \rightarrow q' t$. The s-channel is used in the sub-leading production process, where a virtual $W$ boson is transferred in the s-channel. In this step, the process $q \bar q' \rightarrow t\bar b$ produces a single top quark along with a anti b-quark. The single top quark in combination with a $W$ boson forms up the final state production of the $tW$-channel. A gluon interacts with a b-quark either through the exchange of a virtual top quark or directly in this process. 

The ATLAS experiment is one of the most important LHC experiments for detecting top quark decay products. The top quark decays in $\approx$ 100\% to a $b$ quark and a $W$ boson, according to the measured CKM matrix elements and SM. The b quark hadronizes, forming a narrow cone of hadrons (jet). In the ATLAS detector the exact signature of the top quark depends on the $W$ boson decay mode from the top quark. The $W$ boson can decay leptonicaly ( $\approx$ 32\%) or hadronicaly ( $\approx$ 68\%), with signature of one lepton or two jets respectively and missing energy from a neutrino. In the case of hadronic decay of the top quark's $W$, the top quark is known as hadronic and in the case of leptonic decay of the
top quark's $W$, the top quark is referred as leptonic.

In this work, we studied the studied the applications of deep learning in top pair and single top production at the LHC.

\section{Deep Learning and Artificial Intelligence (AI)}\label{sec2}

Intelligence that has been designed by the humans to accomplish human tasks is known as Artificial Intelligence (AI) and is integrated with computer systems to develop AI tools that eventually works as ``thinking machines" units. It also employs computer science and data to allow machines to solve problems. Some important types of AI is discussed briefly. {\bf Machine Learning (ML)} is kind of artificial intelligence which allows an AI system to automatically adapt from its surroundings and apply that knowledge to make good judgments. It employs a range of algorithms to continuously learn, explain, and enhance data in order to anticipate better results. Computer scientists can use machine learning to `train' a machine by providing it vast volumes of data. To evaluate and make inferences from data, the machine employs a series of rules known as an algorithm. The more data a machine processes, the faster it can perform a task or make a conclusion. ML is further divided into three categories, one is supervised ML, unsupervised ML and semi-supervised ML. ML model requires human involvement when they make a mistake, {\bf Deep Learning (DL)} models can enhance their outputs without the requirement for human involvement through repetition. It involves layers of algorithms and processing units (neurons) into an artificial neural network (ANN). The human brain's structure serves as inspiration for such deep neural networks, where ANN are the elementary unit of deep learning. A Convolutional Neural Network (CNN) is a type of neural network that is capable of handling higher levels of data computation and pre-processing. CNNs were designed especially for image data, and they may be the most adaptable and effective image analysis model .Despite the fact that CNNs were not designed to deal with non-image input, they can produce remarkable outcomes with it.

\subsection{Machine Learning in Experimental High Energy Physics}

The difficulties of identifying small signals from the massive LHC backgrounds have aided the introduction of machine learning (ML) approaches for classification. These methods are a sort of supervised learning in which the outputs are constrained to a small number of variables or classes, such as signals or backgrounds. The data analysis of Higgs boson discovery is a very good example of implication of ML in high energy physics. The CMS~\cite{CMS:2008xjf} and ATLAS~\cite{ATLAS:2008xda} collaborations discovered the Higgs boson in 2012, marking the first time that Boosted Decision Trees (BDT) were used in such a high-stakes investigation for the isolation of small signals (irreducible mass peaks) across huge smoothly dropping backgrounds~\cite{Bourilkov:2019yoi}. 

Another sort of supervised ML is regression algorithms, which produce continuous outputs with any numerical value inside a range. They can be used for reconstruction in high energy physics, for example, where exact measurements of continuous parameters such as track momenta, hit positions or jet energies are required~\cite{Bourilkov:2019yoi}. Machine based categorization methods are quickly replacing conventional HEP procedures for example in jet and particle detection. The LHCb experiment at Large Hadron Collider (LHC) focuses on the physics of `beauty' quarks. The ring-imaging Cherenkov detectors, hadron calorimeters, tracker, electromagnetic calorimeters and muon chambers all require identifying the sorts of long-lived charged particles. Machine learning approaches are being used to develop global particle identification (PID)~\cite{Derkach:2019amb}.

Machine learning is used to detect jets made up of heavy ($c$, $b$, $t$) or light ($u, d, s$) quarks, gluons, and $W$, $Z$, and $H$ bosons in a variety of jet classification challenges. Typically, these categorization issues have been divided into three categories: flavour tagging, which distinguishes between $b, c$ and light quarks; jet substructure tagging, which distinguishes between jets from $W, Z, t$ and $H$; and quark-gluon tagging~\cite{Guest:2018yhq}. The details can be found in two most important classification publications references~\cite{Guest:2016iqz, Baldi:2016fql}.

Quality monitoring is an important step in ensuring that only accurate data, in which all systems work as designed, enter into the physics analysis chain, whereas poor data is identified. At the same time, eliminating useful data is not a wise idea. This takes time at the LHC and includes detectors and online expertise. This is performed in the CMS research by comparing multiple histograms having detector variables to conventional working point references~\cite{Derkach:2019amb}. The CMS collaboration proposes a new method for anomaly identification based on auto-encoders~\cite{Pol:2018fym}. A luminosity section (LS) last for $\approx$ 23 seconds at the LHC and is determined by a specific number of protons orbits in the accelerator. Inside the run, each LS gets its own label, which starts at one and goes up from there. The idea is to use machine learning to give data assurance at the LS level~\cite{Derkach:2019amb}.

Top quark particles play a significant role in a variety of scenarios that are beyond the SM. The top quark have a very high transverse momentum when created by a massive new particle, and will known as boosted top quark. In this phase, studying the large radius jet's substructure, which completely comprises the hadronic decay products, is more effective than resolved reconstruction. Top taggers are the algorithms that detect these decays~\cite{Caudron:2642126}. The decay products of the top quarks are collimated within one big jet at high transverse momenta (\ppt), identifying boosted hadronically decaying top quarks is difficult. Substructure method are used by top tagging algorithms to differentiate
these jets from background jets. In many circumstances, the algorithms use various subjet finding approaches to recreate the jet's decay products. The top quark's mass is another significant variable that can be accessed by different methods. Several algorithms are currently available, with different properties and methods~\cite{Caudron:2642126}.

One of the major purposes of modern machine learning is to extract as much information as possible from a data set. The data structure is used to develop models in successful implementations. Particle collisions in experiments are recreated in high-energy physics (HEP) by mixing the energy deposits left by particles after passing through different segments of a detector. Sub-detector's data can be merged to provide a comprehensive description of each particle generated. Jets are ubiquitous items created in $pp$ collisions at the LHC. Jets are results of the hadronisation of gluons and quarks, which produce a collimated spray of particles. The task of determining the origin of a jet generated by the hadronization of a quark or a gluon is known as tagging. 

Top tagging is a technique used for detecting boosted hadronic top quarks. The general concept for top tagging is to to combine all three jets from a boosted top into single `fat' jet, named as top jet by using the anti-k t jet algorithm with large radius e.g 1.0. Then the fat jets can be tagged to distinguish between background jets and top jets resulting from other processes, by using the internal structure of jets. For example, in p+p collisions, QCD production of two jets. The observable, is one jet shape variable having potential for detecting the top jets, is called N-subjettiness. In a fat jet, it "counts" the number of subjets.

Top and anti-top, which exist in pairs, is the most well-known top-quark production process, and it is based on strong contact. In strong interaction, two strategies for creating top quarks are possible: $gg$ fusion and $q\bar q$ annihilation. Higher order adjustments are done to these processes at NLO, considering both real and virtual gluon emission. Other mixed networks, like $qq$ or $gg$, become available as well.

\section{Data Sets}\label{deeptop}

The public data sets~\cite{Kasieczka:2019dbj, Heimel:2018mkt, Butter:2017cot} are used for current observations which consists of hadronic top signals and backgrounds and mixed quark gluon jets (dijet production) are produced by using pythia~8.2.15~\cite{Sjostrand:2014zea} by ignoring multiple interactions and pile-up at \sqrts~= 14 TeV. In order to include the detector simulation by default ATLAS detector card, Delphes~3.3.2~\cite{deFavereau:2013fsa} has been used. The curved trajectory of charged particles are taken into account by assuming a magnetic field of 2 Tesla and a radius of 1.15 m and also how the tracking efficiency and momentum smearing changes with $\eta$. FastJet~3.1.3~\cite{Cacciari:2011ma, Cacciari:2005hq} is used to define substructure container fat jet through anti $k_t$ algorithm with the $R = 0.8$. In each event only the leading jets are stored having the transverse momentum range: $p_{T,j} = 550 - 650$ GeV.

For signal of jets, the partonic level top quark and its decay product to be in the fat jet ($\Delta R = 0.8$ from the jet axis). No matching is done for the QCD jets also the $|\eta_j| < 2$. Particle flow objects (constituents) extracted from the Delphes Energy-flow algorithm are used as inputs to the subjet analysis, and four momenta of the 200 leading jets are stored. Zero vectors are added for jets having constituents less than 200. They are sorted on the basis of highest {\ppt} on the first. $P_x$ is the truth top 4-momentum, 1 and 0 flags are kept for every jet and is called ``is signal new". Pandas 3.2.2 package of python is use to read the files. They are split into three categories: First is, training contains total 1.2 M events having 600 k signals and background jets each. Second, validation contains total 400 k events having 200 k signals and background jets each. Finally the testing samples contains 400 k events having 200k signals and 200k background jets. All the algorithms are optimised for better comparison using training and validation of data samples, and the results are produced using the test sample.
\begin{figure}[h]
\centering
\includegraphics[width=0.75\textwidth]{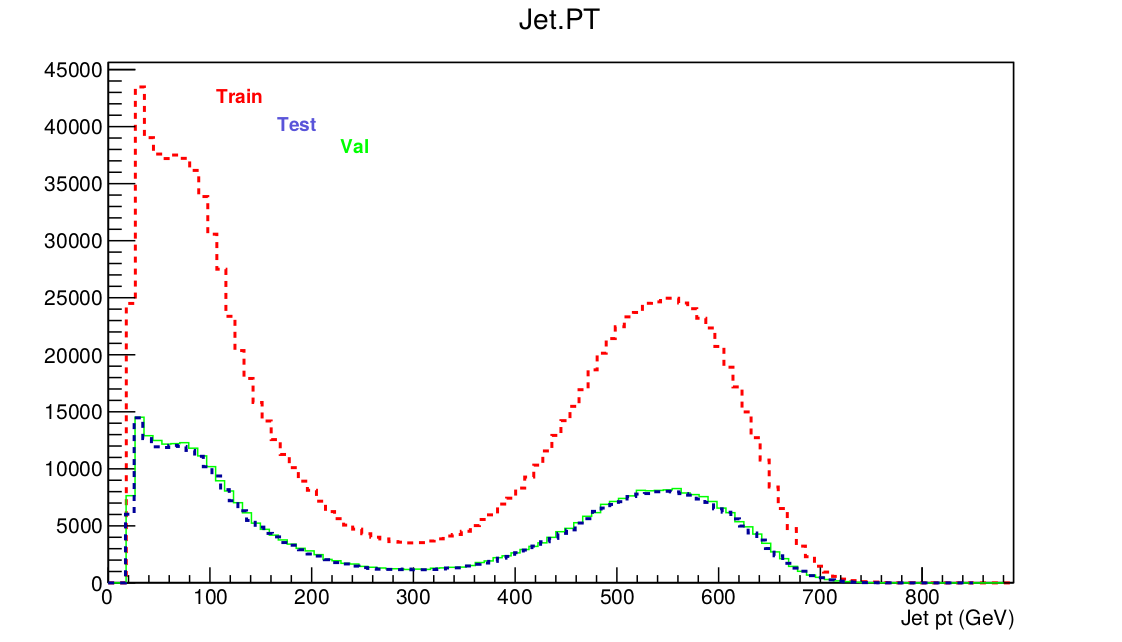}
\caption{Shows the details of the data set containing the test, train and validation samples of the signal events. }
\label{fig1}
\end{figure}

\begin{figure}[H]
     \centering
     \begin{subfigure}[b]{0.49\textwidth}
         \centering
         \includegraphics[width=\textwidth]{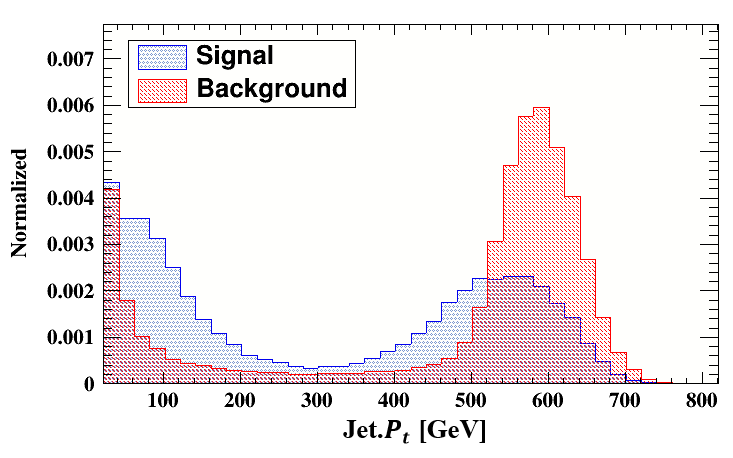}
        \caption{}
         \label{fig2a}
     \end{subfigure}
     \hfill
     \begin{subfigure}[b]{0.49\textwidth}
         \centering
         \includegraphics[width=\textwidth]{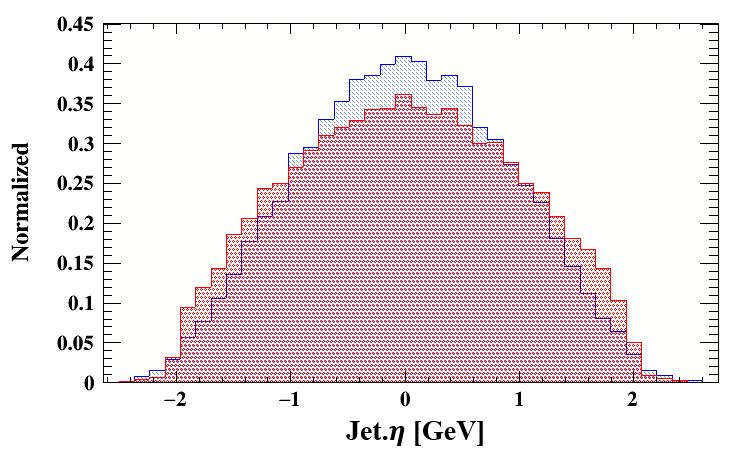}
         \caption{}
         \label{fig2b}
     \end{subfigure}
     \hfill
     \begin{subfigure}[b]{0.49\textwidth}
         \centering
         \includegraphics[width=\textwidth]{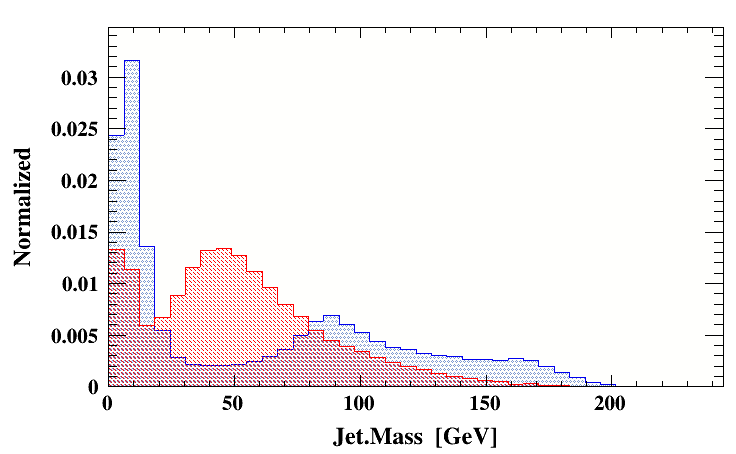}
         \caption{}
         \label{fig2c}
     \end{subfigure}
          \begin{subfigure}[b]{0.49\textwidth}
         \centering
         \includegraphics[width=\textwidth]{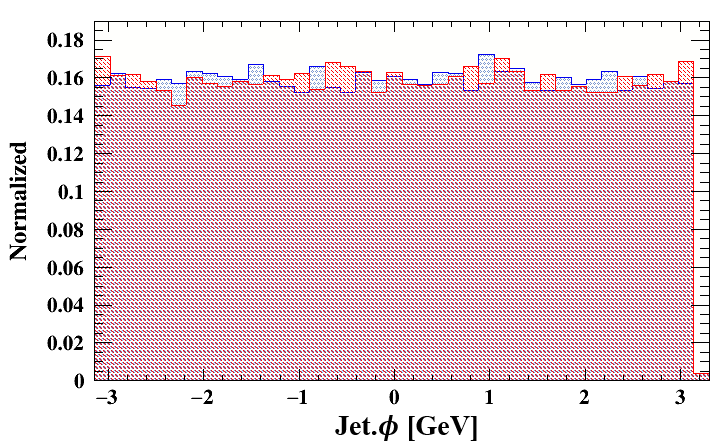}
         \caption{}
         \label{fig2d}
     \end{subfigure}
        \caption{(a) Transverse momentum, (b) pseudo$ \textendash $rapidity ($\eta$), (c) mass and (d) $\phi$ of Jets of the signal and background events}
        \label{fig2abcd}
\end{figure}
\subsection{Model}

A new machine learning approach is used to jet substructure instead of depending on analogies to natural language recognition or image, investigation of the components of the fat jets has been done directly, separating signal from background using only special relativity elements such as the Minkowski metric and Lorentz group. Lorentz layer (LoLa) with a Combination layer (CoLa) is used, and two fully connected layers to generate a deep neural network (DNN) for our DeepTopLoLa tagger. The input 4-moments relate to the calorimeter towers in the standard setup~\cite{Butter:2017cot}. 

\subsubsection{Taggers} \label{subsection1}
A set consisting of $N$ measured 4-vectors are the basic components of any subjet analysis, sorted by {\ppt}. 

\begin{equation*} \label{eq1}
k_{i,j} = 
\begin{pmatrix}
k_{0,1} & k_{1,1} & \cdots & k_{0,N} \\
k_{1,1} & k_{1,2} & \cdots & k_{1,n} \\
k_{2,1} & k_{2,2} & \cdots & k_{2,n} \\
k_{3,1} & k_{3,2} & \cdots & k_{3,n} \\
\end{pmatrix}
\end{equation*}
 
\subsubsection{Combination layer (CoLa)} \label{cola}

Two physics-inspired parts comprises our tagger. We initiate a process by multiplying $C_{l,m}$ with the four vectors from section~\ref{subsection1}. This defines our Combination layer, which is motivated by the treatment of jet clustering in the non-deterministic Q-jets technique

\begin{equation} \label{eq2}
    k_{r,l} \xrightarrow[]{CoLa} \tilde{k}_{r,l} C_{l,m}
\end{equation}

It generates $M$ coupled four vectors $\tilde{k}_m$ from the original $N$ input four-vectors, with and $m = 1...M$ and $l= 1...N$. Our neural network's CoLa matrix has the following trainable form in general.

\begin{equation*} \label{eq3}
C = 
\begin{pmatrix}
1 & 0 & \cdots & 0 & C_{1,N+2} & \cdots & C_{1,M} \\
0 & 1 &  &  \vdots&  C_{2,N+2} & \cdots & C_{2,M} \\
\vdots  & \vdots  & \ddots&0& \vdots & & \vdots \\
0 & 0 & \cdots & 1 & C_{N,N+2} & \cdots & C_{N,M}\\

\end{pmatrix}
\end{equation*}
 
It ensures that the set of $M$ 4-moments $\tilde{k}_m$ contains both a trainable set consisting of linear combinations $M-N$ and the original momentum $\tilde{k}_l$. A deep neural network (DNN) will be used to investigate these $\tilde{k}_m$.

\subsubsection{Lorentz layer (LoLa)} \label{lola}
The relevant distance measured in both two substructure items, according to fundamental theory, is the Minkowski metric. We use it to build a weight function that makes learning the underlying features relatively easy for the DNN. We can choose a conversion that maps the constituent 4-vectors to values more closely related to physical observables because each constituent momentum is described individually by four degrees of freedom. To achieve so, we develop a Lorentz layer as the second portion of the DNN, which converts the $\mathcal{M}$ 4-vectors $\tilde{\mathit{k}}_m$ into the equal number of measurement-motivated items $\hat{\mathit{k}}_m$.
\begin{equation}
\tilde{\mathit{k}}_m\xrightarrow{LoLa}\hat{\mathit{k}}_m=
\begin{pmatrix}
m^{2}(\tilde{\mathit{k}}_m)\\
\mathfrak{p}_{T}(\hat{\mathit{k}}_m)\\
{{w}^{(\mathbb{E})}}_{mn}\mathbb{E}(\tilde{\mathit{k}}_n)\\
{w}^{({\xi})}_{mn}{\xi}^2_{mn}	
\end{pmatrix}
\end{equation}
where ${\xi}^{2}_{mn}$ represent the Minkowski distance between $\tilde{\mathit{k}}_m$ and  $\tilde{\mathit{k}}_m$, 4-momenta.
\[{\xi}^2_{mn}=(\tilde{\mathit{k}}_m-\tilde{\mathit{k}}_n)_{r}{g^{rs}}(\tilde{\mathit{k}}_m-\tilde{\mathit{k}}_n)_{s}\]
and ${w}_{mn}$ is the weights matrices updated all through the network's training. These four entries show various structures that can be used in Lorentz layer. The first  Individual $\tilde{\mathit{k}}_m$ are mapped on invariant mass, and second one is mapped on transverse momentum  by the $\hat{\mathit{k}}_m$. With a trainable vector of weights ${w}^{(\mathbb{E})}_{mn}$, where $n=1,2...\mathcal{M}$, the third entry creates a linear combination of all energies. We can get a set of $\mathcal{M}$ copies of this linear combination by changing the value of m. The fourth entry manages to combine all  $\tilde{\mathit{k}}_n$  with a fixed  $\tilde{\mathit{k}}_m$, along with a trainable weights vector ${w}^{({\xi})}_{mn}$. We enhance the performance of the network by inserting 4 copies with trainable weights that are independent for the last entry with the weights ${w}^{({\xi})}_{mn}$. From these 4-copies, two minimize over internal index and remaining two sum over it.
The network model i.e. LoLa is trained using KERAS with THEANO and ADAM optimizer. The leaning rate is 0.001. The learning rate determines the step size in the numerical minimization of loss-function. The training of the model terminates, when the performance of test sample does not improve after several epochs. After training of the model with initial seed weights, the performance is then tested by the validation sample.

\begin{figure}[H]
     \centering
     \begin{subfigure}[b]{0.49\textwidth}
         \centering
         \includegraphics[width=\textwidth]{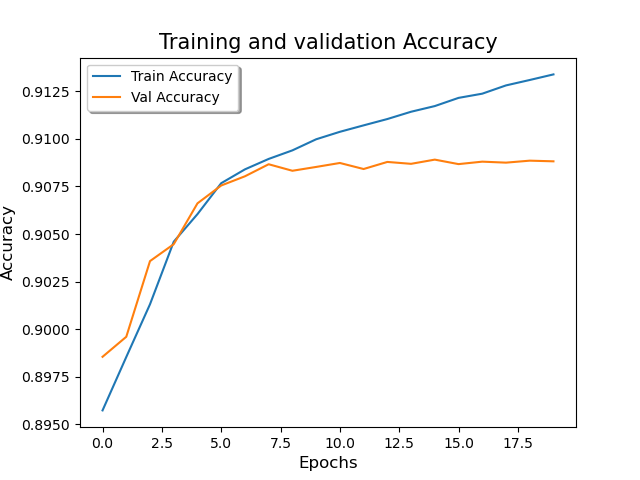}
         \caption{}
         \label{fig3a}
     \end{subfigure}
     \hfill
     \begin{subfigure}[b]{0.49\textwidth}
         \centering
         \includegraphics[width=\textwidth]{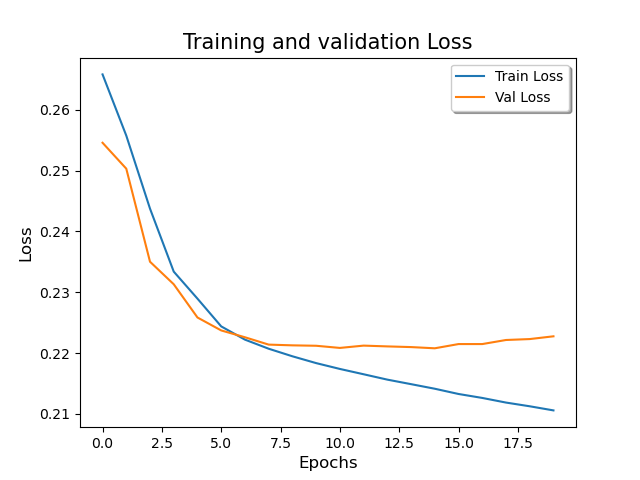}
         \caption{}
         \label{fig3b}
     \end{subfigure}
     \hfill
     \begin{subfigure}[b]{0.49\textwidth}
         \centering
         \includegraphics[width=\textwidth]{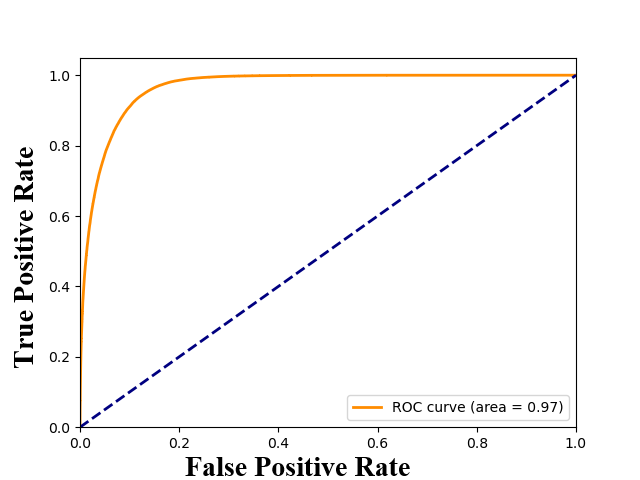}
         \caption{}
         \label{fig3c}
     \end{subfigure}
        \caption{ (a) Epoch vs Accuracy, (b) Epoch vs Loss, (c) ROC curve for the tagger DeepTopLoLa }
                \label{fig3abc}
\end{figure}

        Figure 3(a) shows the epoch as a function of accuracy curve, which approaches 90\% and shows the efficiency of our model. Its classification accuracy trained to recognize the signal (top quark) form the backgrounds. Figure 3(b) shows the training and validation loss versus number of epochs, that shows how after each epoch loss is reducing. Training terminates after 17 epoch and shows how after each epoch loss is reducing. Figure 3(c) is Receiver Operating Characteristic curve (ROC curve), False positive rate (FPR) vs True Positive Rate (TPR). As area under the curve (AUC) is 0.97 so we conclude from ROC curve that used model~\ref{lola} method have high classification accuracy of separating signal from backgrounds. These plots are drawn for LoLa algorithm, so similarly we can obtain plots for other algorithms by making change in our main file. LoLa is for Learning with Opponent-Learning Awareness, a system where each entity modifies the expected learning of other entities in the environment.

\section{Multivariate Analysis For Jet substructure Study}
TMVA stands for Toolkit for Multivariate Data Analysis with the ROOT framework. In HEP, to explore for the smaller signals from huge data sets, maximum information has to be extracted for analysis. Based on machine learning, the Multivariate classification technique becomes basic for many of the data analysis methods. Incorporated with the ROOT framework, the TMVA toolkit has a huge diversity of classification algorithms. Training, testing, progress assessment and use of all accessible classifiers are executed at the same time and are simple to use. Supervised Machine Learning is used in all TMVA methods. They utilize the training events which determines the required outputs. TMVA is mainly planned for the requirements of HEP applications, however ought not be limited to these~\cite{Hocker:2007ht}.

The TMVA classification and regression analysis are mainly classified into two independent stages: Firstly, the training stage where the testing, training and evaluation has been done for the multivariate methods governed by TMVA factory and second is the application stage where the picked methods have been applied by using TMVA Reader to specific regression or classification problems that were trained before. 

\subsection{Data Pre-processing, Training and Testing}

The data sets needs to be provided to the factory that are required for training and testing. The classifiers are trained and tested using data sets in the form of ROOT that contain known event classification. In this scenario, the factory divides the tree into two sections: one for training and another for testing. When provided in the data collection, individual event weights can be assigned. The connection between user and the TMVA analysis processes is organized by the factory class, which includes pre-analysis and pre-processing of the data needs to be trained to determine basic features of the variables used as an input to the classifiers. Once the training has completed, each classifier stores its transformation in it's weight file. The weight file includes all of the information needed for the afterwards application of the trained classifier. The transformation is taken from its weight file for testing and application of the classifier. The classifiers are tested and evaluated once they have been trained in order to determine their performance.

In machine learning, a classifier is an algorithm which automatically sorts or classifies data within one or more 'classes'. Email classifier is the most common example, which examines emails and filters them according to whether they are spam or not. A classifier is an algorithm - the principles that machines use to categorise data. In contrast, the classification model is the final outcome of a classifier's machine learning. The classifier is used to train the model, and the model is then used to classify your data. The signal is overlaid by background processes with the same signature in the high energy. Widely used techniques of classification into signal and background events meet their limits, whenever the signal is very weak, or when a piece of information about whether an event is signal or background is hidden and not very well established correlations between the observables. The Multivariate Analysis Toolkit TMVA allows you to efficiently extract the information from the observables.

In TMVA, classification is performed by generating a classifier output from the input variables (observables) in which signal events have values equivalent to 1 and background values appraoches to 0. The mapping to the background class and signal is carried out by labelling all events as signal with a classifier output $y > y_{cut}$ and remaining are classified as background. Purity of the signal efficiency $\epsilon_{sig, eff}$ and background rejection $(1-\epsilon_{bkg, eff})$ are measured for each cut value $y_{cut}$~\cite{Speckmayer:2010zz}. The light-weight Reader object is used to apply the trained classifiers to the chosen events from a data sample with signal and background composition. It could be involved in any C++ executable, ROOT files, or python analysis program and reads and interprets the weight files of the specified classifier~\cite{Voss:2007jxm}.
\subsection{Methodology}
The {\bf boosting} strategy has proven to be a very successful method of enhancing performance not only for decision trees, but also for any type of classifier. In high-energy physics, boosting, and particularly boosted decision trees, has grown increasingly popular~\cite{Coadou:2013lca}. The D0 experiment uses boosted decision trees for the first time to conduct a search, and it resulted in the first indication (and then observation) of single top quark production~\cite{D0:2006ngk}. A decision tree takes a collection of input features and divides input data depending on those features in a recursive manner. They only evaluate signal and background trees because they only consider trees having two classifications. A decision tree's initial node is the root node. Until a stopping condition is reached, each node can be split into two daughters or we can say branches~\cite{Coadou:2013lca}. The phase space is divided into numerous areas that are finally designated as signal or background based on the number of training events that wind up in the final leaf node. In a regression tree, each output node represents a different value of the target variable. The trees are constructed from the same training ensemble by reweighting events, and then combined into a single classifier (regressor) that is average of a individual decisions (regression) trees. Boosting stabilizes the behavior of decision trees regarding variations in the training sample and can significantly improve performance when compared to a single tree~\cite{Voss:2007jxm}.

Consider the problem of distinguishing between signal and background activities using a set of particle identification (PID) variables. The decision tree is a set of binary data divisions. The tree is trained using a collection of well-known training events. The findings are measured using a distinct set of known testing events. Assume that all data is stored on a single node. The best PID variable to classify the data into signal and background is discovered. There are two nodes after that. The operation is continued on these new nodes until a particular number of end nodes (referred to as ``leaves") is produced, or until all of the leaves are pure, or until a node has fewer events~\cite{Roe:2005hm}.

There are many widely used criteria for determining the appropriate PID variable and the best location for splitting a node. We used gini criterion for boosted decision trees. Assume that $W_j$ is the weight of event $j$. The term purity $P$ of node can be defined as, the ratio of the weight of signal events $W_s$ on the node and total weight ($W_s + W_b$) of events on that node. i.e
\begin{equation} \label{eq4}
    P = \frac{W_s}{W_j + W_j} 
\end{equation}
gini is defined for a specific node as:

\begin{equation} \label{eq5}
    gini = P (1- P \sum_{j} W_j)
\end{equation}
where, the value of $gini$ is 0 for both $P$ = 0 and 1. The split that minimises $gini_{left} + gini_{right}$ is preferred as the best. The next node to split is determined by identifying which node's splitting maximises $gini$ change. A decision tree is created
in this manner. Signal leaves have value of P greater than or equal to 0.5, while background leaves have value P less than 0.5. Decision trees are significant, but they are also unstable. A minor change in the training data can lead to a large shift in the tree. Boosting is used to solve this problem. The AdaBoost is a widely used boosting classifier. An exponential loss function is used in adaptive boost. The weights of the training events that were misclassified are raised (boosted) for boosting, and a new tree is built. Then for the new tree, the procedure is repeated. Many trees are formed in this manner.
\begin{figure}[H]
\centering
\includegraphics[width=0.70\textwidth]{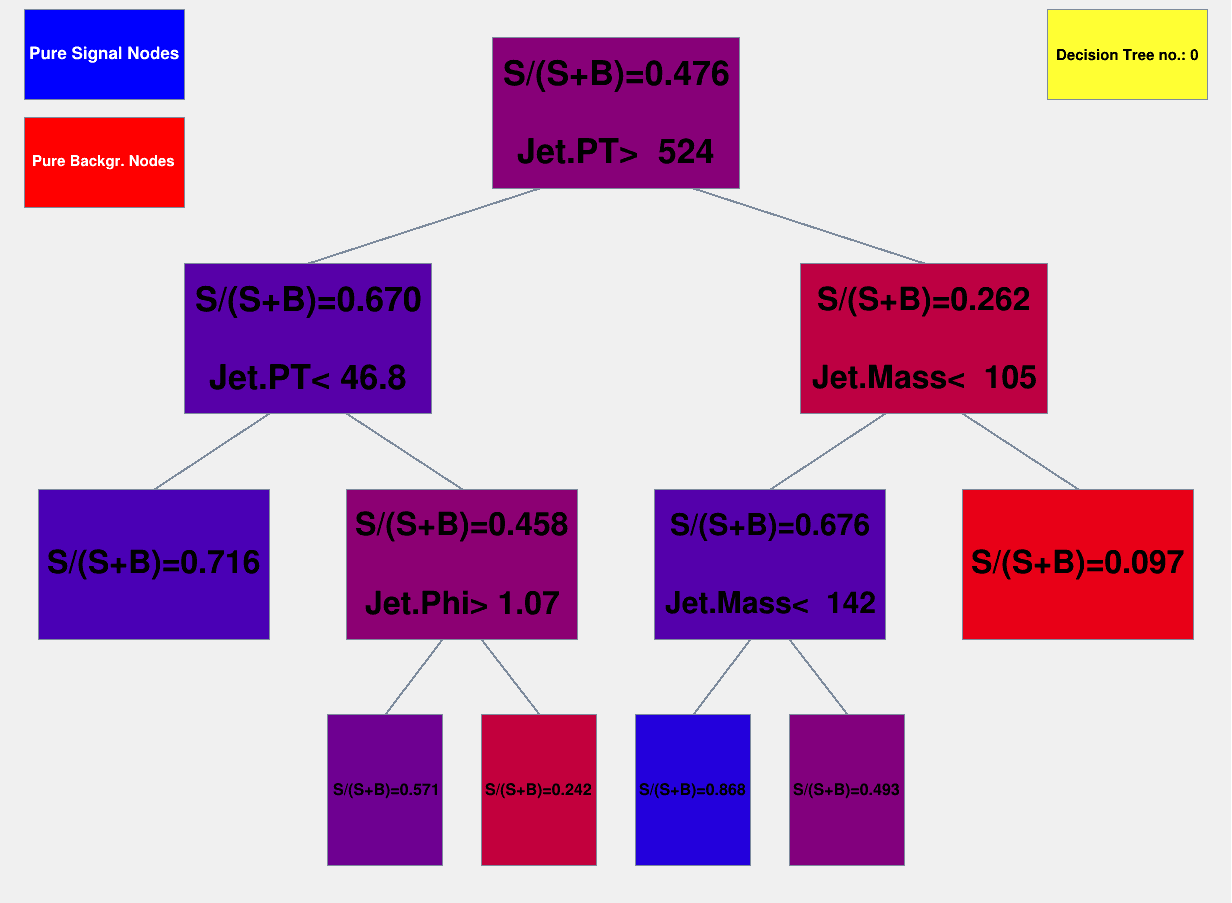}

\caption{Sepration of signal and background of initial tree of BDT with adaptive booster by applying selection cuts of $Jet.p_T>20$}
\label{fig4}
\end{figure}

If the event occurs on a signal leaf, then the score $T_n$ from the nth individual tree is +1, and if the event occurs on a background leaf, then score is -1. The final score $T$ is calculated as a weighted sum of the individual leaf scores $T_n$.

\begin{equation} \label{eq6}
    T = \sum_{n=1}^{N_{tree}} \alpha_n T_n
\end{equation}
where,
\begin{equation}\label{eq7}
    \alpha_n = \beta log \frac{1-err_n}{err_n}
\end{equation}
and here $err_m$ represents the error function (misclassification rate) for the nth tree, given by
\begin{equation}\label{eq8}
    err_n = \frac{weight~misclassified}{total~weight~for~tree~m}
\end{equation}
and $\beta$ is constant its typically used values are 1 0r 1/2. In a decision tree, the dividing criteria for every node are defined by the process of training, building, or growing the tree. The training process begins with root node, which determines an initial dividing criterion for the entire training sample. When the training events are separated, they are divided into two subsets, each of which is subjected to the same algorithm to decide the next dividing iteration. We need to repeat this process till the entire tree is constructed. At every node, the splitting is determined by calculating the variable and related cut value that offers the best signal-to-background separation. When the least number of events set in the BDT configuration is achieved, the node splitting stops (option nEventsMin). On the basis of how many events are present, the leaf nodes are categorised as signal and background nodes. In theory, the splitting might go on till every leaf node only includes signal or background events, indicating that complete classification is possible. Such a decision tree, on the other hand, would be severely over trained. A decision tree must have been pruned to prevent over training.
Although a neural network with 'n' number of neurons can theoretically have $n^2$ directional links, the complexity can be minimised by layering the neurons and permitting only direct links from one layer to the next. The \textbf{multi-layer perceptron} is the name given to this type of neural network; in TMVA all neural network implementation  are of this type.The input layer of a multilayer perceptron is the first layer, the output layer is the last, and the rest are hidden layers. In a classification problem with $n_{var}$ input variables, the input layer contains $n_{var}$ neurons that keep the input values, while the output layer contains one neuron that keeps the output variable.

In Multilayer Perceptron (MLP), the data is transferred from input layer to output layer in a forward direction, similar to a feed forward neural network. The neurons of the MLP are trained using the back propagation learning algorithm (a technique for optimising the weights of an MLP that use the outputs as inputs). We used default TMVA configuration of MLP having 600 number of epoch (NCyles), Hidden Layers = $N+5$ with $n_{var} -1$ neurons and $tanh$ activation function.

The maximum \textbf{likelihood} method involves in creating a model out of probability density functions (PDFs) that regenerates the signal and background input variables.The likelihood of an event that is of signal type is calculated by multiplying the signal probability densities of all input variables, that are considered to be independent (Naive Bayes), and normalising by the total of the signal and background likelihoods for a given event. This PDE method is also known as the "naive Bayes estimator" because correlations between variables are neglected.
The Naive Bayes is a probability based algorithm used in machine learning technique that may be applied to a wide range of classification tasks. For event $j$ the likelihood ratio $y_L(j)$ is defined as follows:
\begin{equation}\label{eq8}
    y_L(j) = \frac{L_s (j)}{L_s (j)+L_B (j)}
\end{equation}
here
\[L_{S(B)}(j)=\prod_{i}^{n_{var}}\mathfrak{p}_{S(B),i}(z_{i}(j))\]
the background and signal PDFs for the ith input variable $z_i$ are $\mathfrak{p}_{B,i}$ and $\mathfrak{p}_{S,i}$ respectively. The PDFs have been normalised as,
\begin{equation*}
\int_{-\infty}^{+\infty}\mathfrak{p}_{S(B),i}(z_{i})dz_i=1\,\,\,\, \forall{i}
\end{equation*}
In absence of model inaccuracies (that is correlations among input variables that were not eliminated by the de-correlation method, or a probability density model that was inaccurate), it can be demonstrated that, For the given set of input variables, the ratio~\ref{eq8} gives the perfect signal obtained from background splitting. As an outcome of TMVA, the likelihood output can be transformed using an inverse sigmoid function that zooms in with the peaks.
\subsection{Deep-Top}
For the training purpose, two types of trees are involved: the signal tree and the background tree. The signal tree consists of all the features describing the signal events. The background tree consists of exactly the same features passed through the same preselection cuts as the signal. The distributions for these input variables for deeptop datasets '\ref{deeptop}' are shown in figure\ref{fig4} and the network architecture used for analysis is shown in Fig.~\ref{fig6}.
\begin{figure}[H]
\centering
\includegraphics[width=0.75\textwidth]{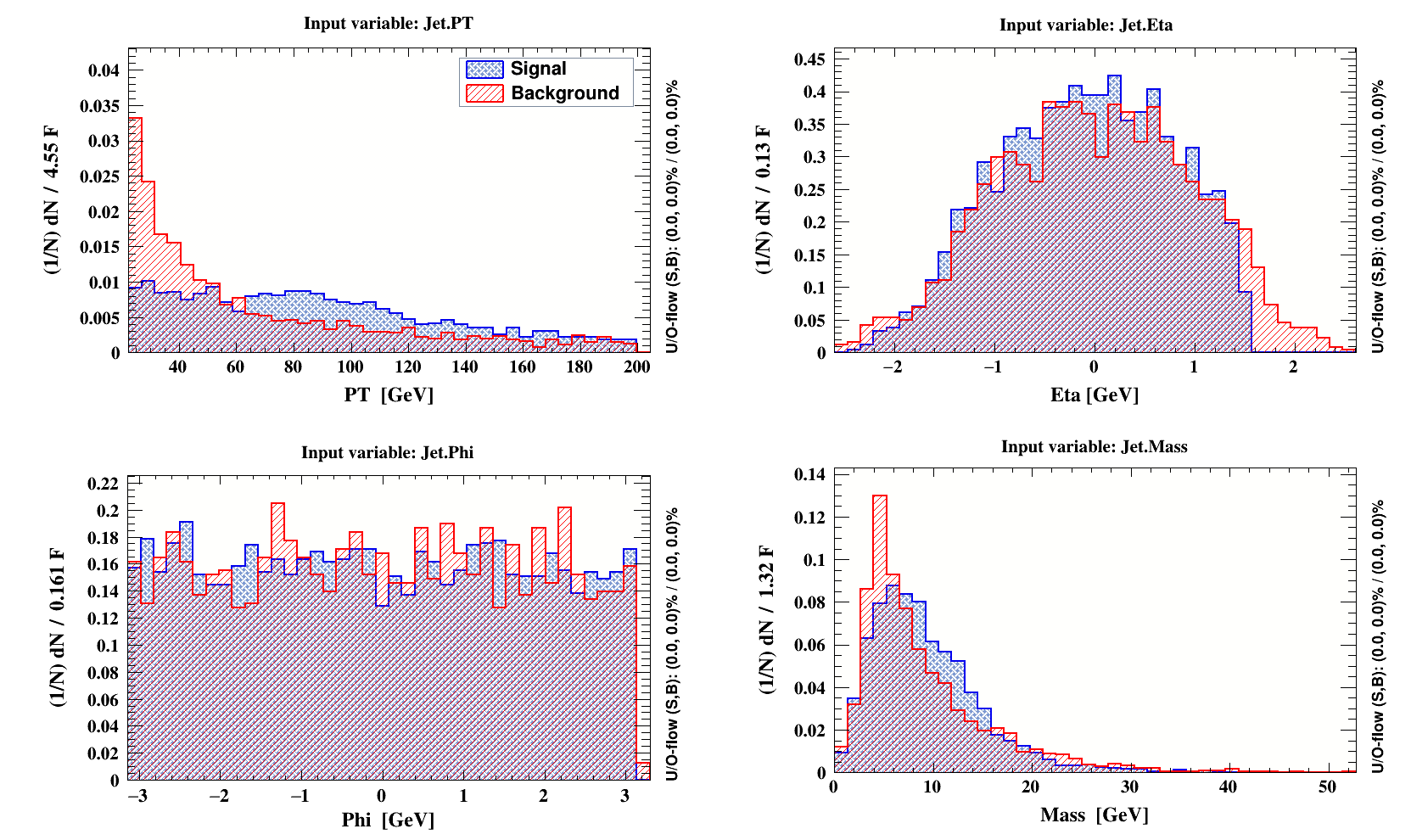}
\caption{Input variables of Deep-Top for multivariate analysis.}
\label{fig5}
\end{figure}
\begin{figure}[H]
\centering
\includegraphics[width=0.60\textwidth]{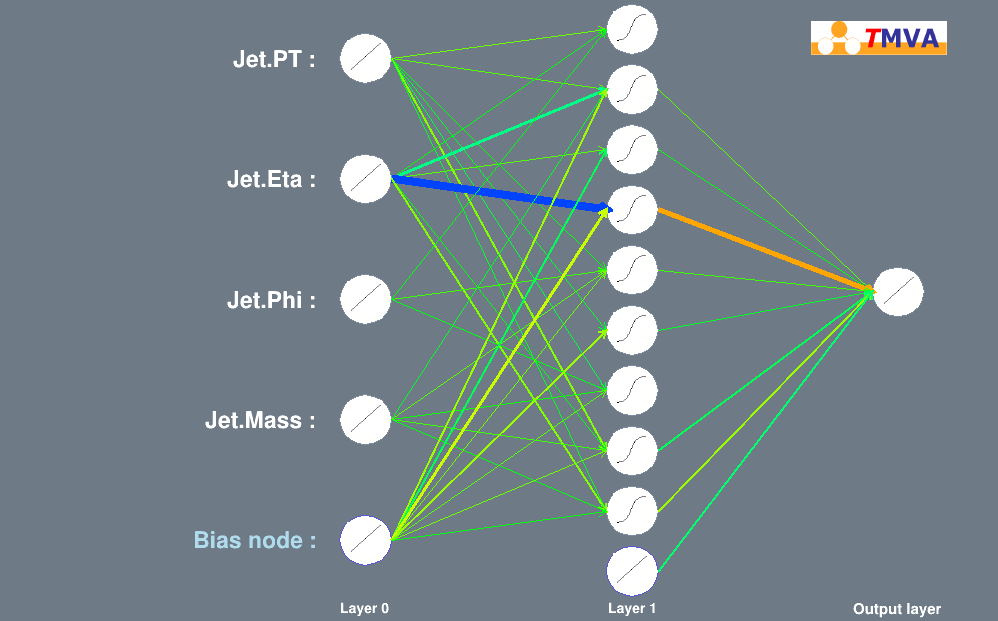}
\caption{Network Architecture for the multivariate analysis.}
\label{fig6}
\end{figure}
\subsubsection{Results}
\begin{figure}[h]
     \centering
     \begin{subfigure}[b]{0.4\textwidth}
         \centering
         \includegraphics[width=\textwidth]{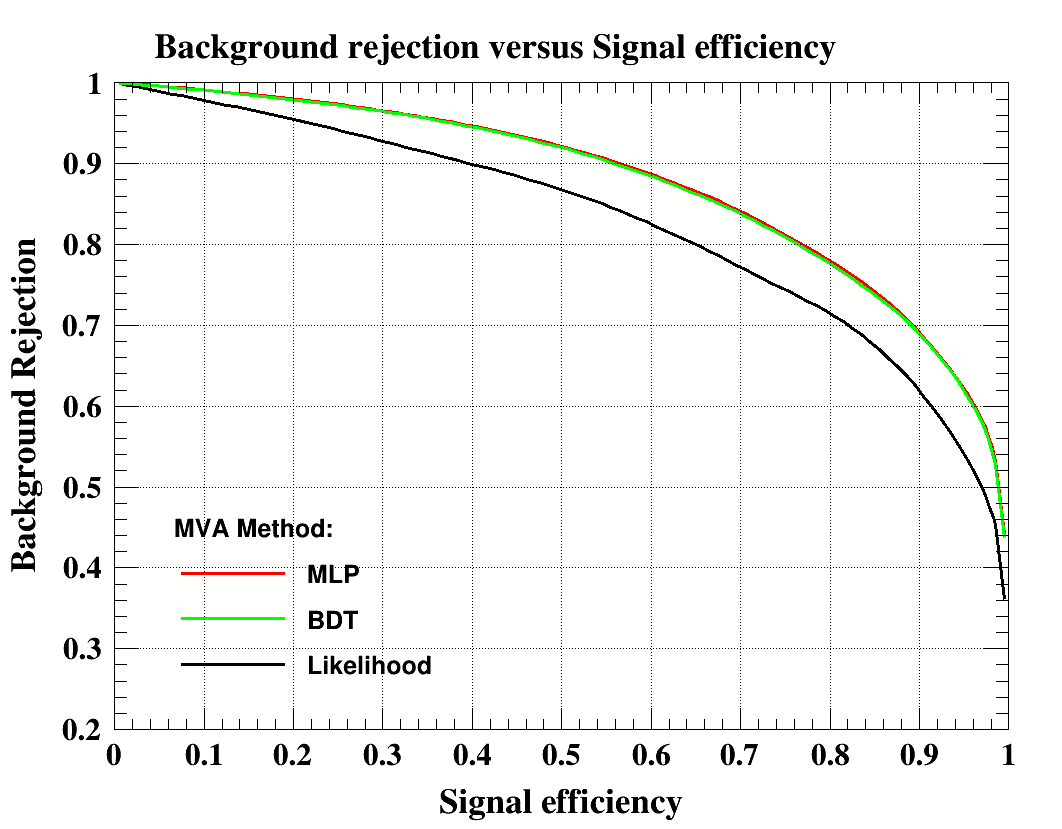}
         \caption{}
         \label{fig7}
     \end{subfigure}
     \hfill
     \begin{subfigure}[b]{0.45\textwidth}
         \centering
         \includegraphics[width=\textwidth]{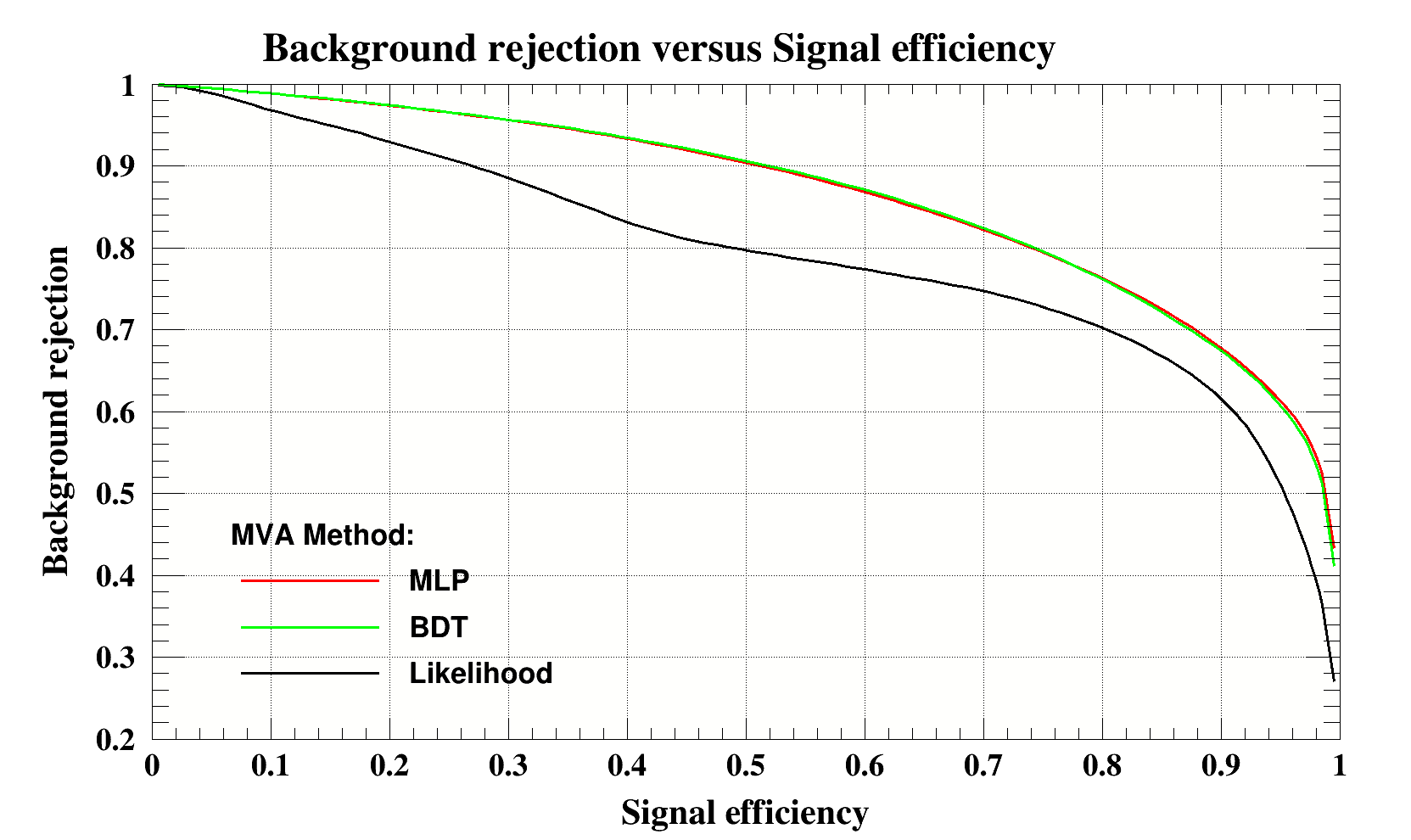}
         \caption{}
         \label{fig8}
     \end{subfigure}
     \hfill
       \caption{(a) and (b) represents Background$ \textendash $Rejection Vs Signal$ \textendash $Efficiency with and without applying cuts respectively.}
        \label{fig7.8}
\end{figure}

The curve of background$ \textendash $rejection against signal$ \textendash $efficiency provides a reasonable estimate of a classifier's performance. A high background rejection and a high signal efficiency are required to achieve the best separation of signal and background candidates. A classifier's performance is measured by the area under the background$ \textendash $rejection vs. signal$ \textendash $efficiency curve, which is also used to rank the classifier. The bigger the area, the better a classifier's predicted separation power. The order of the classifiers in the fig. \ref{fig6} and fig. \ref{fig7} corresponds to their rank with applying cut and without applying cut respectively, with the best performing classifier appearing first. 
\begin{table}[h]
	\begin{center}
		\begin{tabular}{||c c c||} 
			\hline
			MVA Classifier & AUC (with cut) & AUC (without cut) \\
			\hline\hline
			MLP &  0.877 & 0.865\\ 
			\hline
			BDT  & 0.875 & 0.864 \\
			\hline
			Likelihood  & 0.829 & 0.793\\
			\hline
		\end{tabular}
		\label{table:1}
		\caption{MVA Classifier Area Under the Curve with and without cut value}
	\end{center}
\end{table}

\begin{figure}[h]
     \centering
     \begin{subfigure}[b]{0.45\textwidth}
         \centering
         \includegraphics[width=\textwidth]{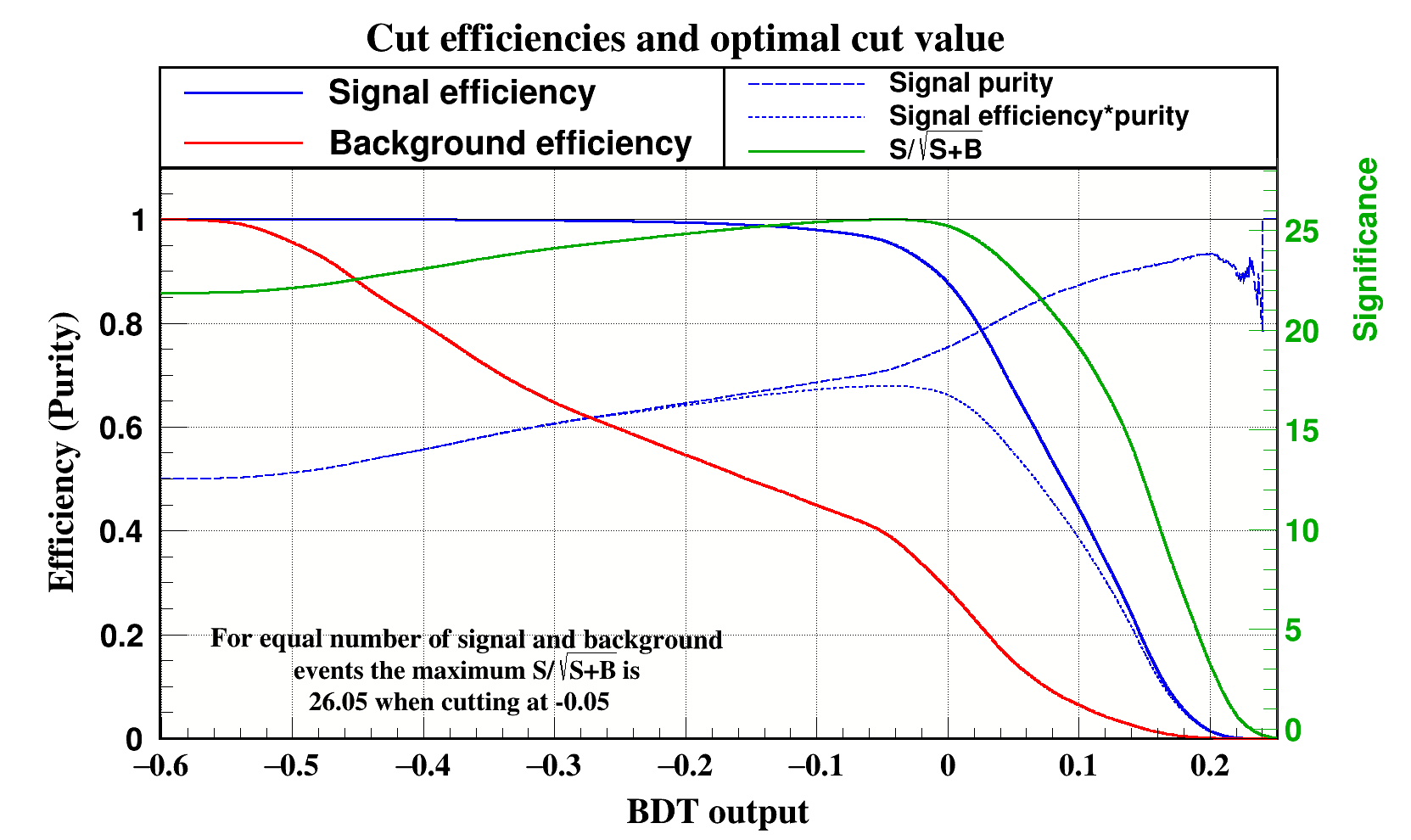}
         \caption{}
         \label{fig9}
     \end{subfigure}
     \hfill
     \begin{subfigure}[b]{0.45\textwidth}
         \centering
         \includegraphics[width=\textwidth]{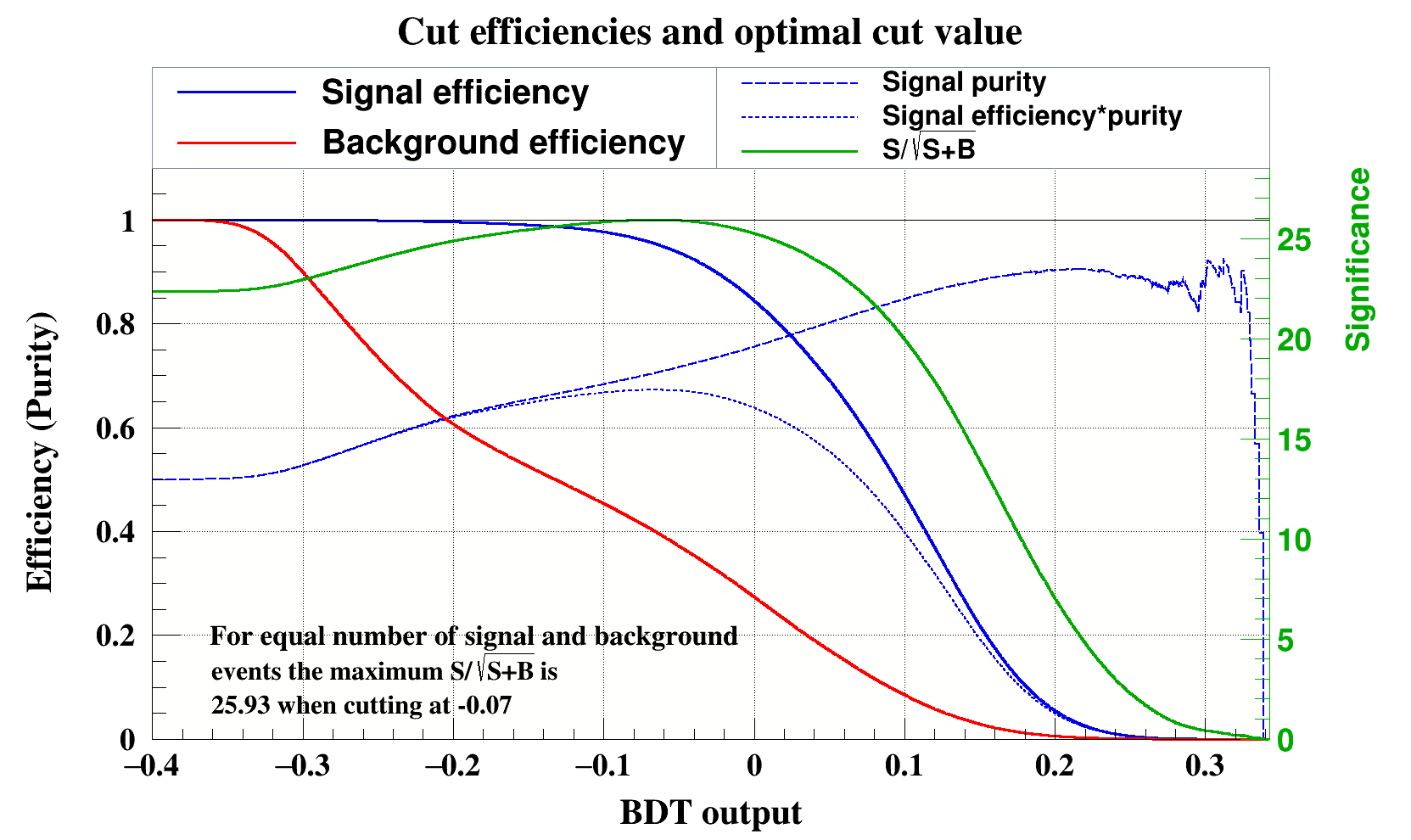}
         \caption{}
         \label{fig10}
     \end{subfigure}
     \hfill
       \caption{(a) and (b) represents BDT Signal Significance with and without applying cuts respectively.}
        \label{fig9,10}
\end{figure}

\begin{figure}[h]
     \centering
     \begin{subfigure}[b]{0.45\textwidth}
         \centering
         \includegraphics[width=\textwidth]{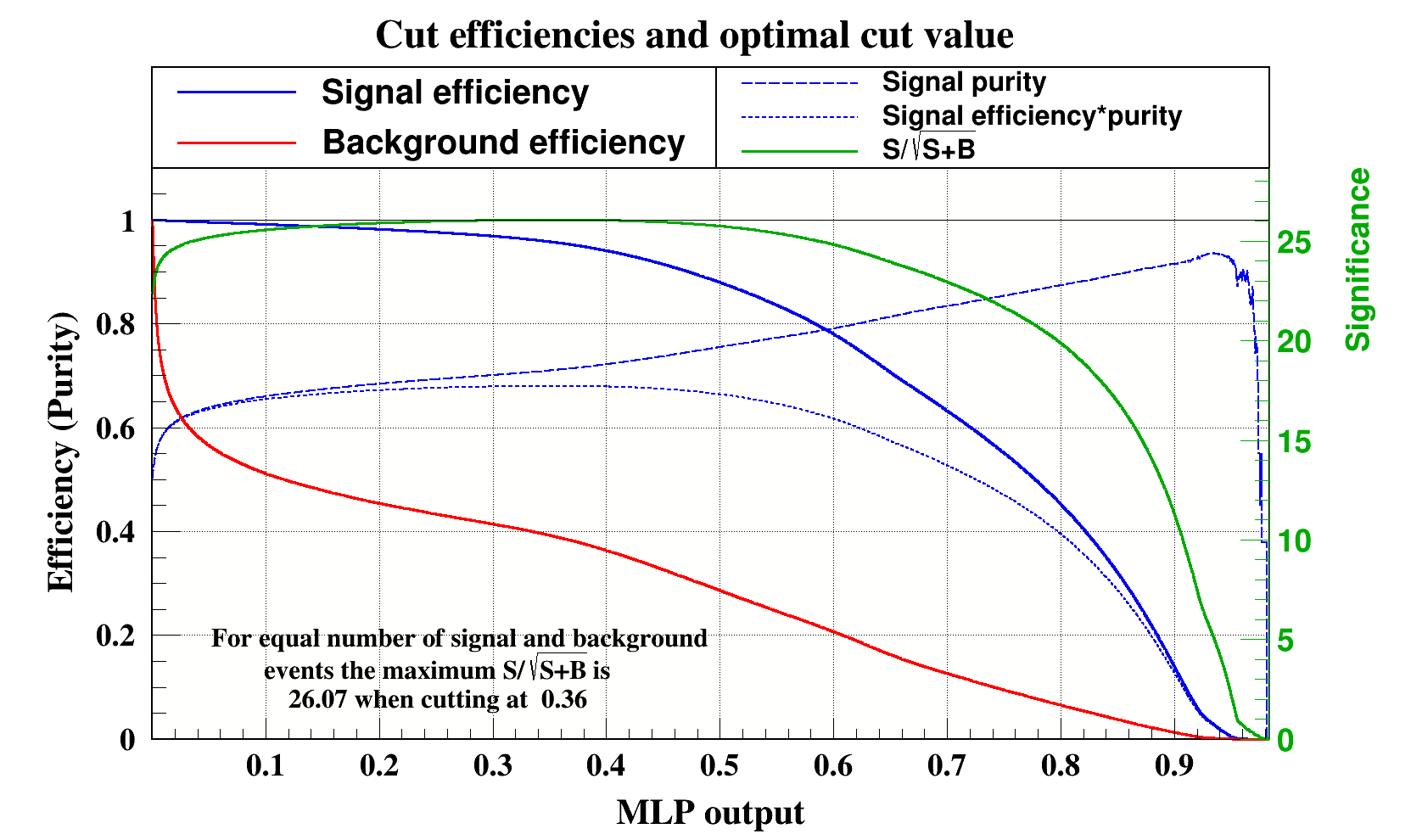}
         \caption{}
         \label{fig11}
     \end{subfigure}
     \hfill
     \begin{subfigure}[b]{0.45\textwidth}
         \centering
         \includegraphics[width=\textwidth]{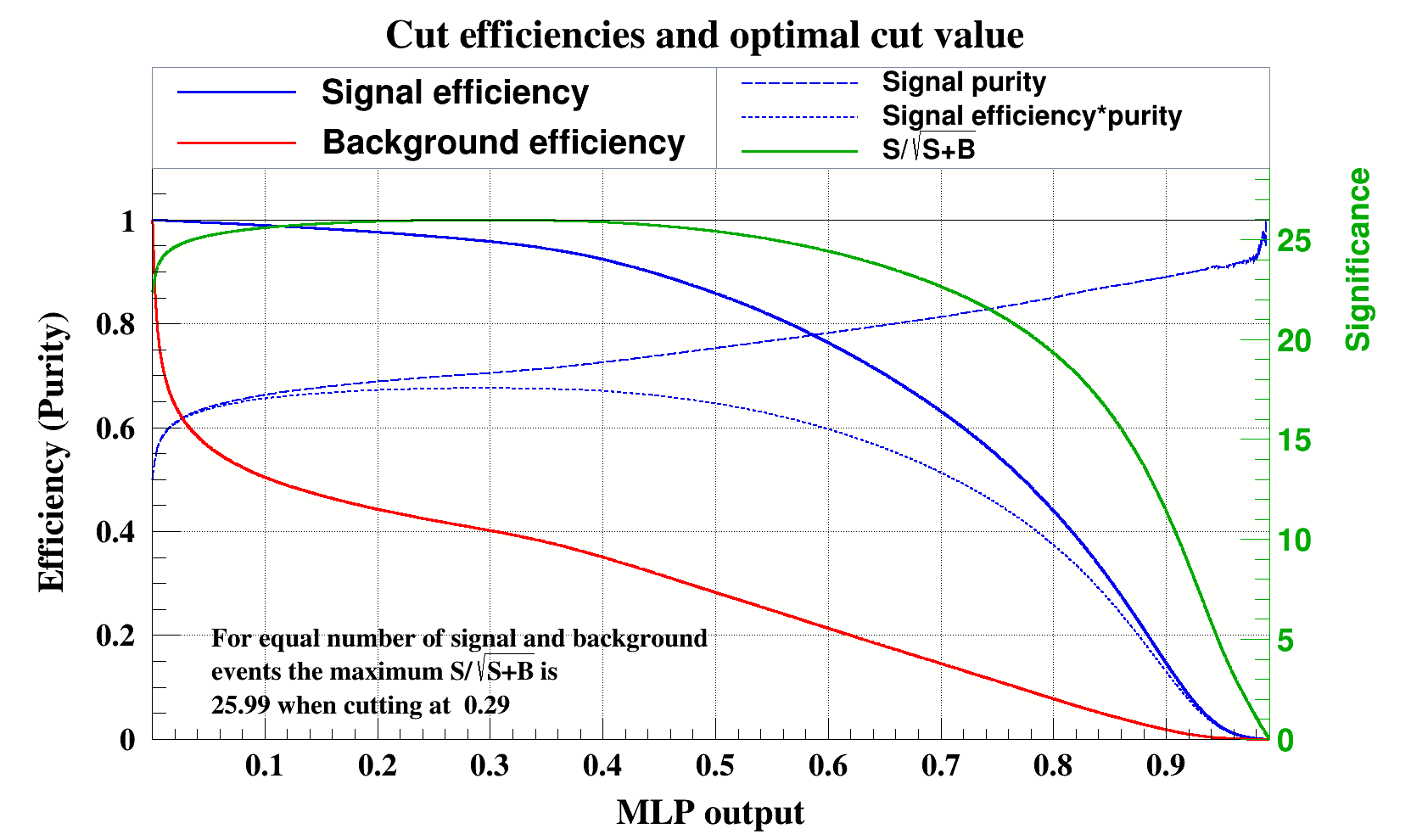}
         \caption{}
         \label{fig12}
     \end{subfigure}
     \hfill
       \caption{(a) and (b) represents MLP Signal Significance with and without applying cuts respectively.}
        \label{fig11,12}
\end{figure}

\begin{figure}[h]
     \centering
     \begin{subfigure}[b]{0.45\textwidth}
         \centering
         \includegraphics[width=\textwidth]{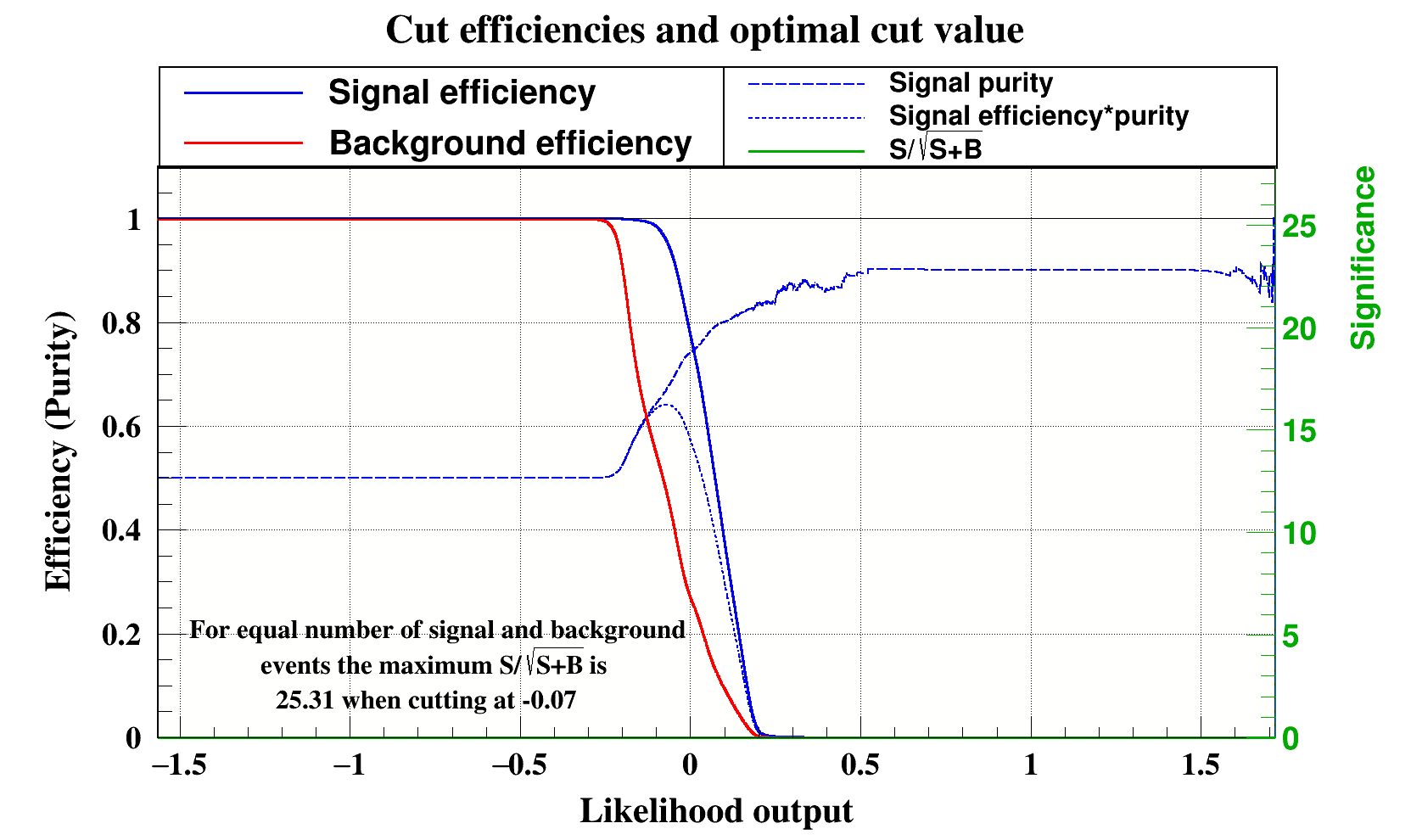}
         \caption{}
         \label{fig13}
     \end{subfigure}
     \hfill
     \begin{subfigure}[b]{0.45\textwidth}
         \centering
         \includegraphics[width=\textwidth]{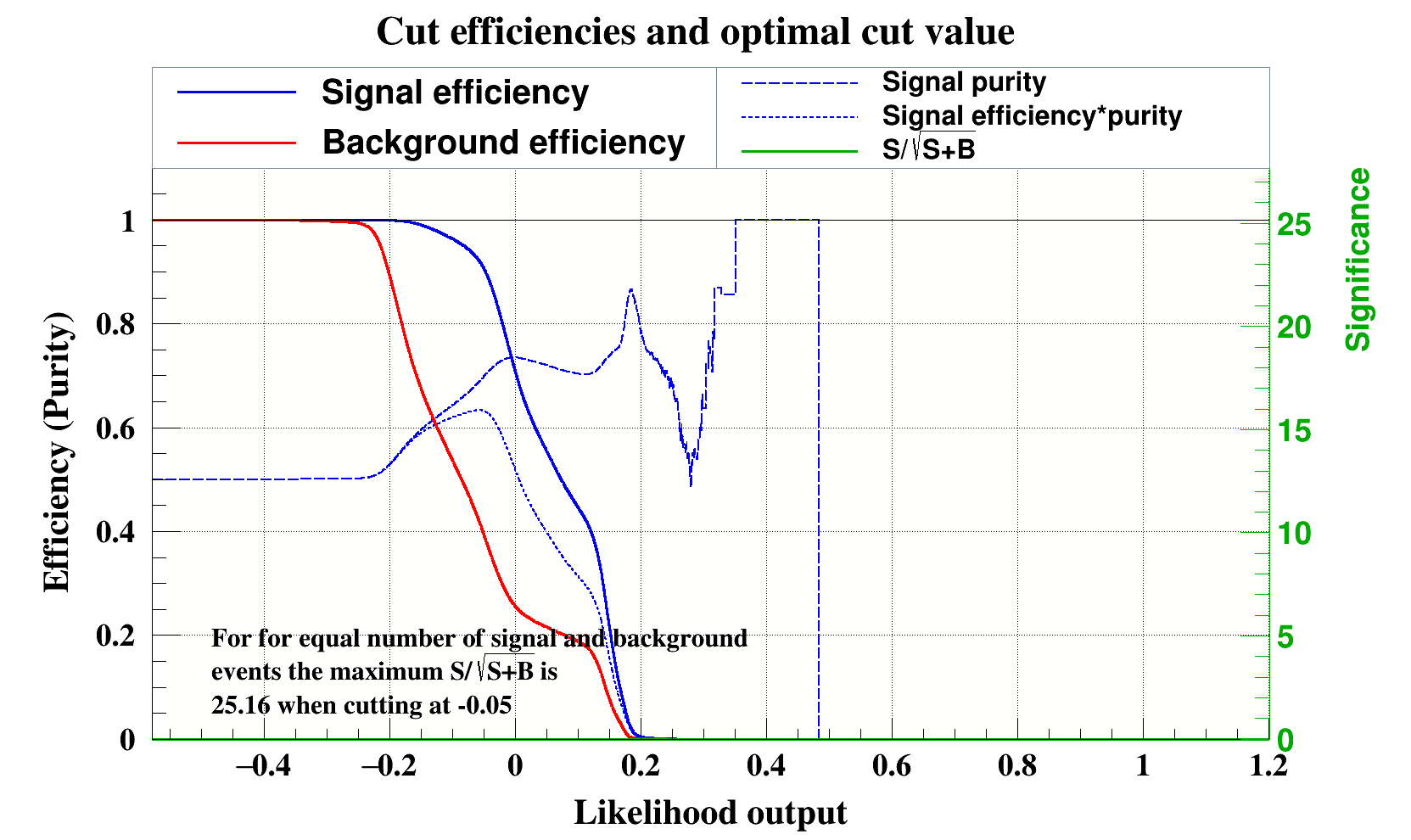}
         \caption{}
         \label{fig14}
     \end{subfigure}
     \hfill
       \caption{(a) and (b) represents likelihood  Signal Significance with and without applying cuts respectively.}
        \label{fig13,14}
\end{figure}

Each classifier is trained for 50000 number of signal and background events. To check the visibility of the signal process, signal$ \textendash $significance is calculated for each classifier. Signal$ \textendash $significance is calculated by using a cut on the jet$  P_{t} $ as follow:
\begin{center}
	$ Jet_{P_{t}} > 20 $ GeV
\end{center}
Figures \ref{fig9} and \ref{fig10} shows the comparison of signal$ \textendash $significance for BDT classifier with and without applying cut value respectively. It is clear that the signal significance increases by applying cut on  jet$P_{t}. $\\
Similarly signal significance is calculated for MLP and Likelihood classifiers with and without applying cut value on jet$P_{t}$ as shown in fig. \ref{fig11} to fig. \ref{fig14}.

The comparison of signal significance for each classifier is shown in table 2:
\begin{table}[H]
	\begin{center}
		\begin{tabular}{||c c c||} 
			\hline
			MVA Classifier & Signal Sig. (with cut) & Signal Sig. (without cut) \\
			\hline\hline
			MLP  & 26.07 & 25.99\\ 
			\hline
			 BDT &  26.05 & 25.93\\
			\hline
			Likelihood  & 25.31 & 25.16\\
			\hline
		\end{tabular}
		\label{table:2}
		\caption{Signal Significance with and without cut value}
	\end{center}
\end{table}

\begin{figure}[h]
     \centering
     \begin{subfigure}[b]{0.45\textwidth}
         \centering
         \includegraphics[width=\textwidth]{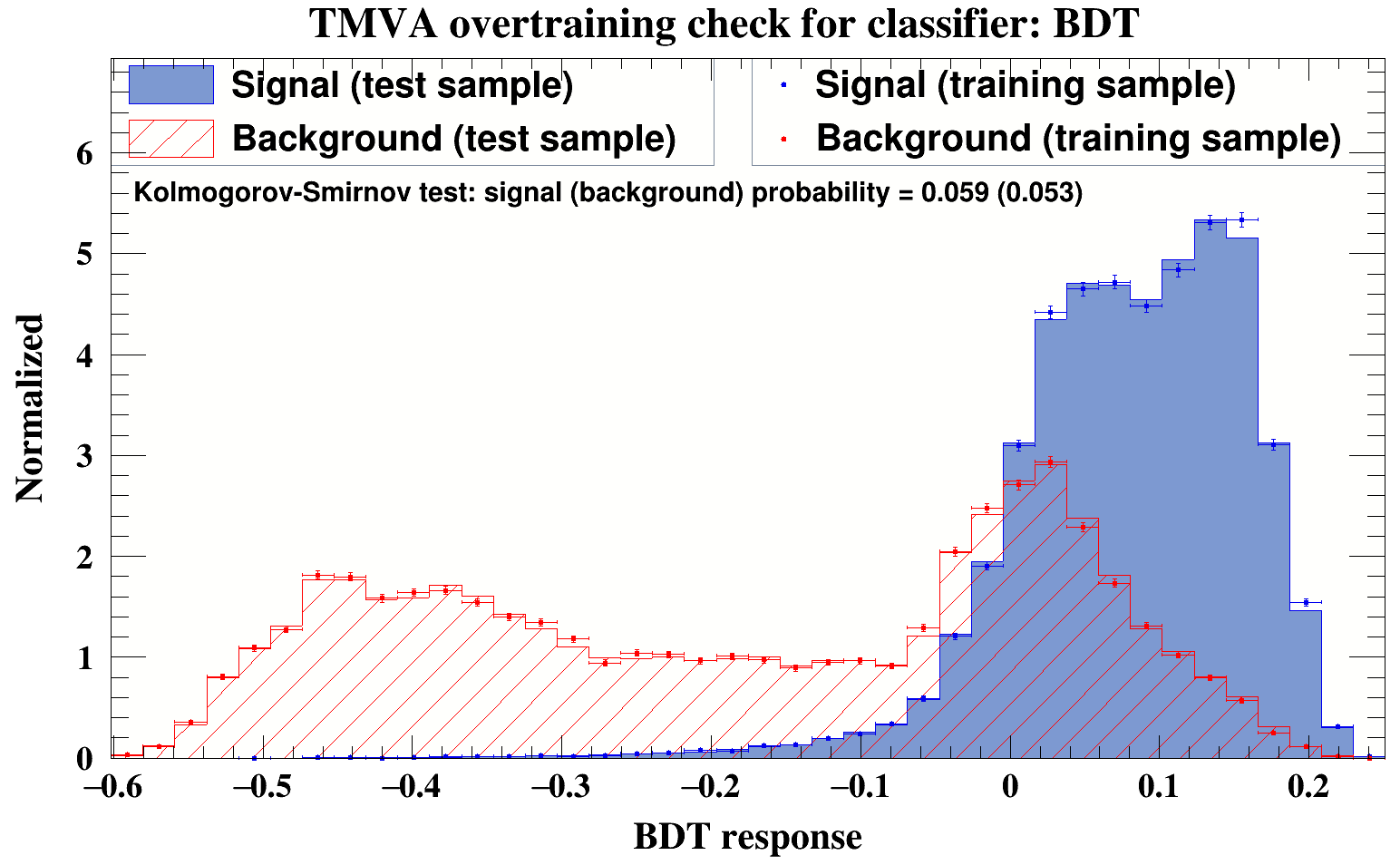}
         \caption{}
         \label{fig15}
     \end{subfigure}
     \hfill
     \begin{subfigure}[b]{0.45\textwidth}
         \centering
         \includegraphics[width=\textwidth]{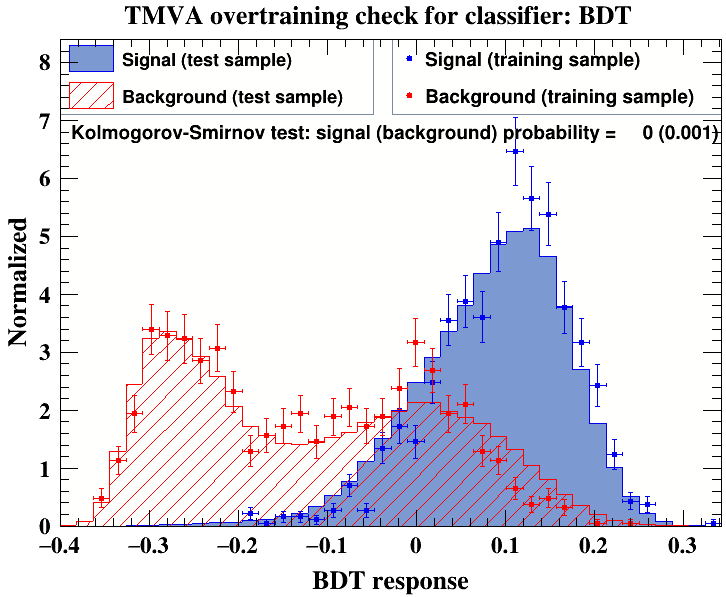}
         \caption{}
         \label{fig16}
     \end{subfigure}
     \hfill
       \caption{(a) and (b) represents BDT response with and without applying cuts respectively.}

{(a) and (b) represents BDT response with and without applying cuts respectively.}
        \label{fig15,16}
\end{figure}

\begin{figure}[h]
     \centering
     \begin{subfigure}[b]{0.45\textwidth}
         \centering
         \includegraphics[width=\textwidth]{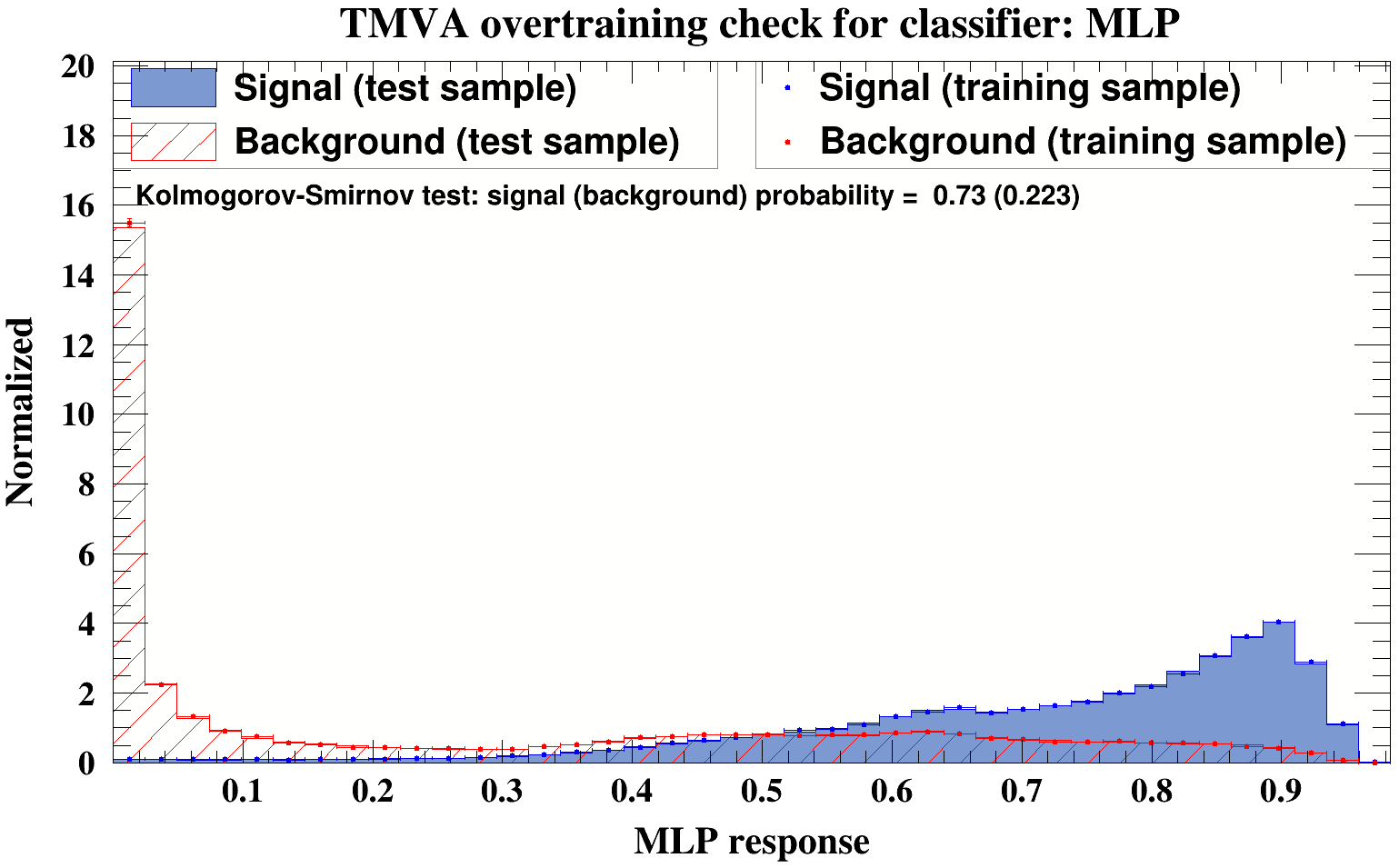}
         \caption{}
         \label{fig17}
     \end{subfigure}
     \hfill
     \begin{subfigure}[b]{0.45\textwidth}
         \centering
         \includegraphics[width=\textwidth]{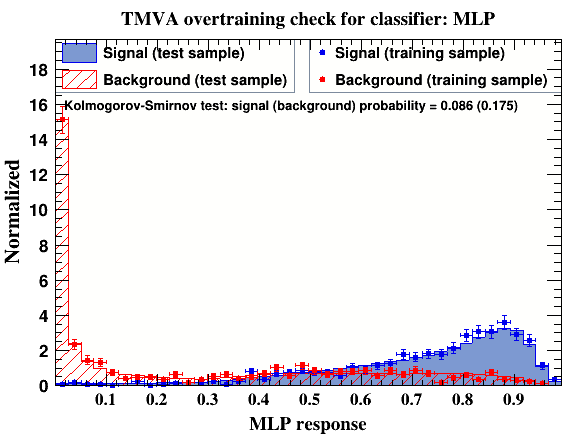}
         \caption{}
         \label{fig18}
     \end{subfigure}
     \hfill
       \caption{(a) and (b) represents MLP response with and without applying cuts respectively.}
        \label{fig17,18}
\end{figure}

Classifier's response gives signal to background probability by using Kolmogorov-Smirnov (KS-Test). For the BDT classifier signal (background) probability is almost zero without applying cut, but it increases if a cut is applied on $ jet P_{t} $ as shown in Figures \ref{fig15} and \ref{fig16} respectively. The over-training decreases and the performance of the classifier improve by applying cut. \\Similar behavior can be observe in the MLP classifier as shown in Figure \ref{fig17} and Figure \ref{fig18}.

\begin{figure}[h]
     \centering
     \begin{subfigure}[b]{0.45\textwidth}
         \centering
         \includegraphics[width=\textwidth]{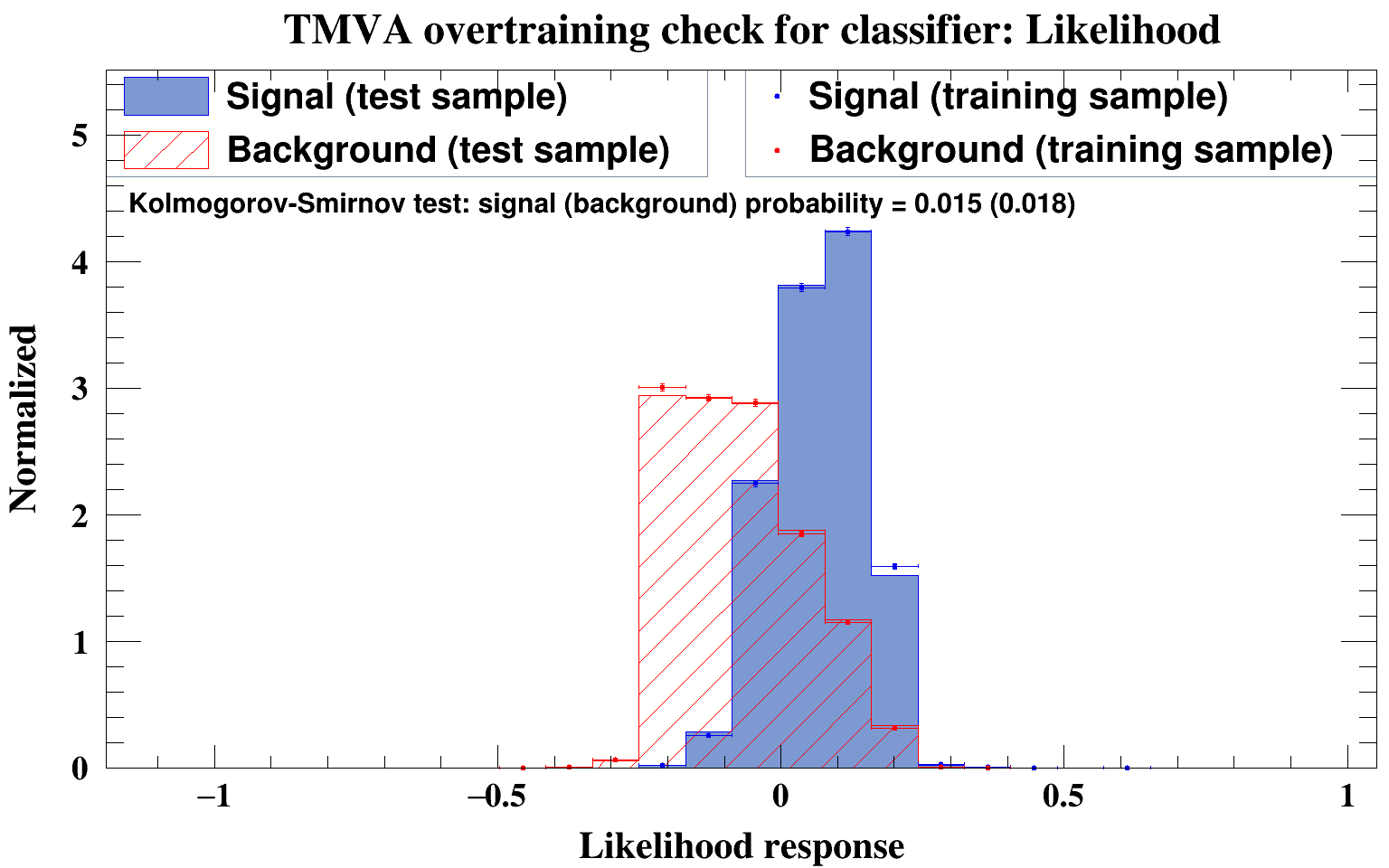}
         \caption{}
         \label{fig19}
     \end{subfigure}
     \hfill
     \begin{subfigure}[b]{0.45\textwidth}
         \centering
         \includegraphics[width=\textwidth]{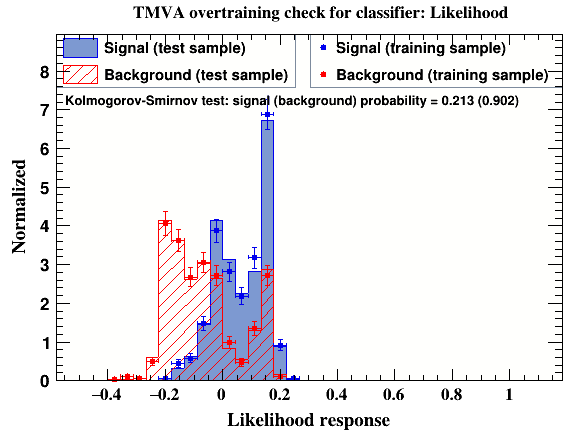}
         \caption{}
         \label{fig20}
     \end{subfigure}
     \hfill
       \caption{(a) and (b) represents MLP response with and without applying cuts respectively.}
        \label{fig19,20}
\end{figure}

But for the liklihood the classifier performance decreases and over-training increases by applying cuts which can be seen in the Figure \ref{fig19} and Figure \ref{fig20}.

\begin{table}[H]
	\begin{center}
		\begin{tabular}{||c c c||} 
			\hline
			MVA  & Signal (Background) & Signal (Background) \\
			\hline
			Classifier & (with Cut) & (without Cut)\\
			\hline     

			\hline
			MLP  & 0.73(0.223) & 0.086(0.175)\\ 
			\hline
			BDT &  0.059(0.053) & 0(0.001)\\
			\hline
			Likelihood  & 0.015(0.018) & 0.213(0.902)\\
			\hline
		\end{tabular}
		\label{table:3}
		\caption{Comparison of classifier's response with and without cut value}
	\end{center}
\end{table}
For the same sample of data, three classifiers i.e. BDT, MLP and Likelihood are trained. Background rejection, signal efficiency, signal significance and classifier response is computed for each classifier with and without $ JetP_{t} $ cut. The comparison of computed results shows that among all three classifiers Multi-layer Perception (MLP) has the best output for all the calculated parameters.

\subsection{Multivariate Analysis for Single Top Quark}
Our second kind of datasets belong to single top quark production and the relevant background processes for TMVA has been created through Calchep~3.1.1 by using `Standard Model' with the 7 Tev energy of each colliding particle and resulted output in the `lhe' format which is then converted into root by Delphes~3.1.2 using the ATLAS detector card. The data consists of 100 k events of T-channel as signal tree and background Trees comprising of S-channel excluding $W$'s, S-channel excluding $H$'s (using 2HDM-II-BSM model) and $t \bar t$. Each tree consists of 100 k events. The input variables we used for training are transverse momentum ({\ppt}) of jet(Jet.{\ppt}), angle $\phi$ and $\eta$ (pseudorapidity) of jet and last one is mass of jet.

\begin{figure}[H]
\centering
\includegraphics[width=0.60\textwidth]{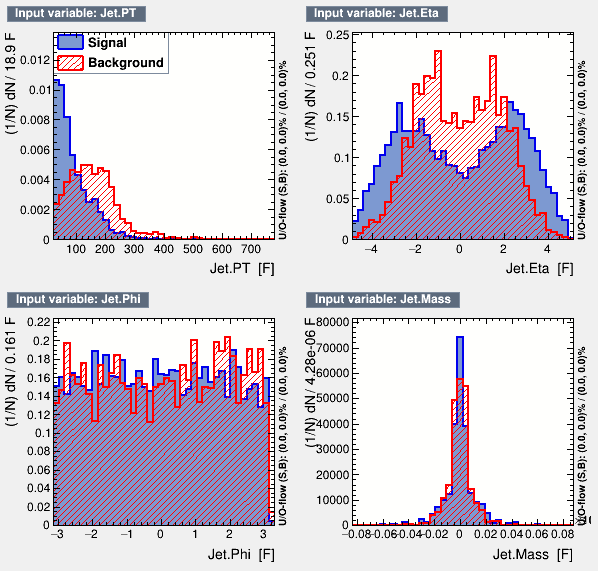}

\caption{Input variables distribution for signals and backgrounds. }
\label{fig21}
\end{figure}

\begin{figure}[H]
\centering
\includegraphics[width=0.60\textwidth]{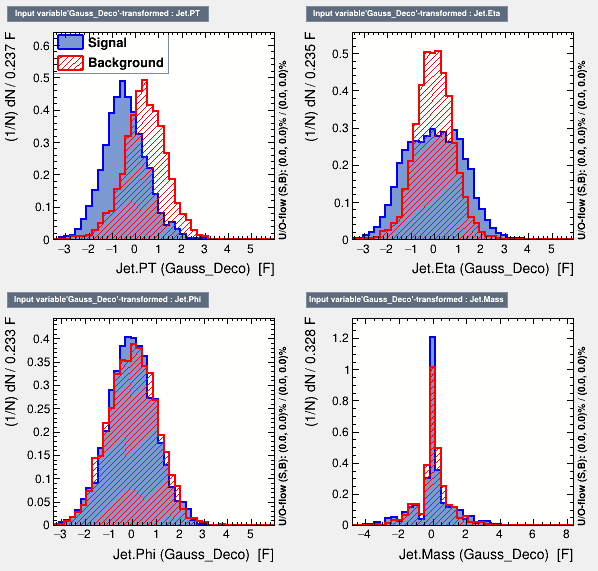}

\caption{Input variables distribution for signals and backgrounds after application of the Gaussian Transformation. }
\label{fig:fig22}
\end{figure}

The input variables we used for training are jet (Jet.{\ppt}), angle $\phi$ and $\eta$ and last one is mass of jet as shown in figure~\ref{fig12}. We observe the peak at almost 50 $GeV$ for signals and 200 $GeV$ for backgrounds with selection cut of {\ppt} greater than 20 $GeV$. The backgrounds tends to give larger {\ppt} than the signals. 177,396 signal events passed and 577,556 number of events passed of backgrounds, out of which 5000 events from each are used for training and remaining for testing. After applying the gaussian distribution, the input variables distribution for the signal and background can be seen in Fig.\ref{fig:fig22}.

\section{Results and discussion}\label{sec3}

There are several multivariate analysis methods but currently we choose three of them. For each technique, we use the default TMVA setting. We choose 5000 events for each to train the signals and backgrounds. To enhance the performance of the trained BDT, we used 850 trees, with node splitting permitted only when the number of events in a node exceeds 2.5\% of the total number of events in the training sample. The maximum tree depth has been set at three. Adaptive Boost is used for training, with a learning rate of $\beta= 0.5$. At the end of each boosting iteration, half of the training sample is chosen at random. The separation index of the parent node and the sum of the indices of the two daughter nodes are compared to optimise the cut value on the variable in a node. For the separation index, we use the Gini Index. Finally, the variable's entire range is evenly gridded into 20 cells.

\begin{figure}[H]
\centering
\includegraphics[width=0.4\textwidth]{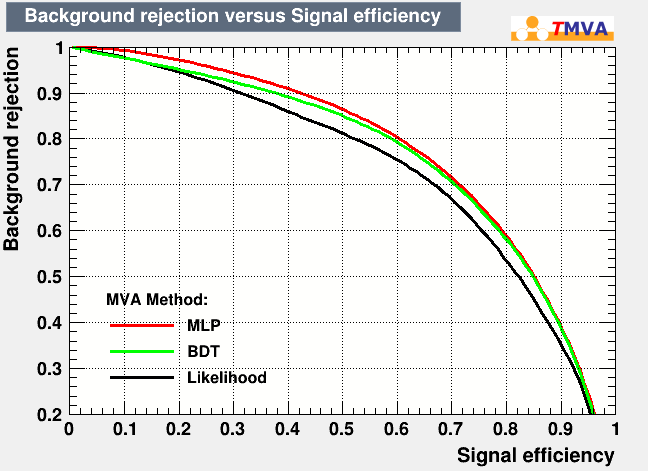}

\caption{Receiver Operating Characteristic (ROC) curves from TMVA for several multivariate analysis. This represents signals efficiency($\epsilon_{sig, eff}$) and background rejection ($1-\epsilon_{sig, background}$). }
\label{fig23}
\end{figure}

Figure~\ref{fig23} shows the receiver operating characteristic (ROC) curves for the selected methods. ROC curves indicates the quality of the training for the booked multivariate methods. The ROC curve plots the rate of false positive (background rejection) against the rate of true positive (signal efficiency). The true positive rate (sensitivity) refers to the number of true signal events that have been identified correctly. The false positive rate (1-specificity), on the other hand, is a measure of how many actual background events have been correctly identified.The larger the area under the curve, better the performance. It has been observed that the multilayer Perceptrons (MLP) shows better efficiency and background rejection as compared to other methods. BDT performance is in between MLP and likelihood methods. Area Under the Curve (AUC) of ROC is $0.761$, $0.773$ and $0.734$ for BDT, MLP and Likelihood methods respectively.

\begin{figure}[h!]
\centering
\includegraphics[width=0.4\textwidth]{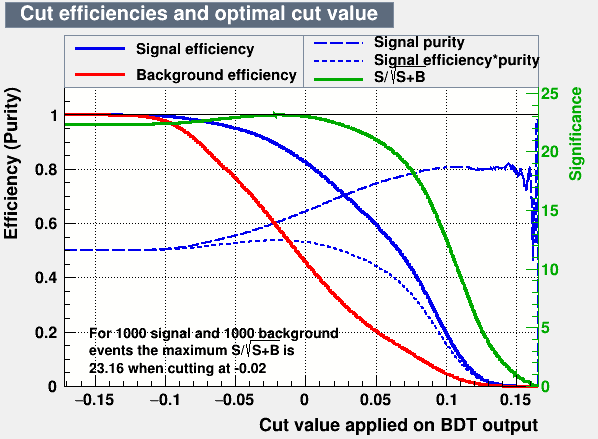}

\caption{Signal efficiency, background efficiency, signal purity, signal efficiency
multiplied with purity and significance $\frac{S} {\sqrt{\mathrm{S+B}}}$ as function of the BDT cut value.}
\label{fig24}
\end{figure}
For a given number of signal and background events, figure \ref{fig24} depict the significance $\sqrt{\mathrm{S+B}}$ as a function of the BDT cut value. This is the parameter that most properly indicates the BDT's training performance. The significance output is used to evaluate a method's performance , with a higher significance indicating a better classifier. However, statistics must be considered, which means that the signal efficiency that corresponds to the ideal cut value must be considered. A lot of signal gets filtered away if this number is really low. If the real data has low statistics, the quality of the cut cannot be guaranteed because the output distribution differs from that of the training sample.

\begin{figure}[H]
     \centering
     \begin{subfigure}[b]{0.4\textwidth}
         \centering
         \includegraphics[width=\textwidth]{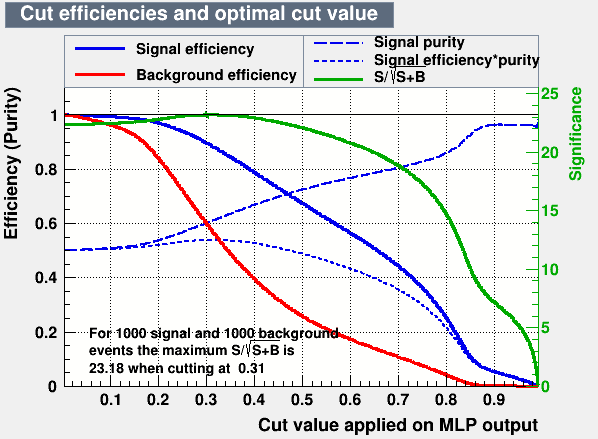}
         \caption{SMLP Response.}
         \label{fig25a}
     \end{subfigure}
     \hfill
     \begin{subfigure}[b]{0.4\textwidth}
         \centering
         \includegraphics[width=\textwidth]{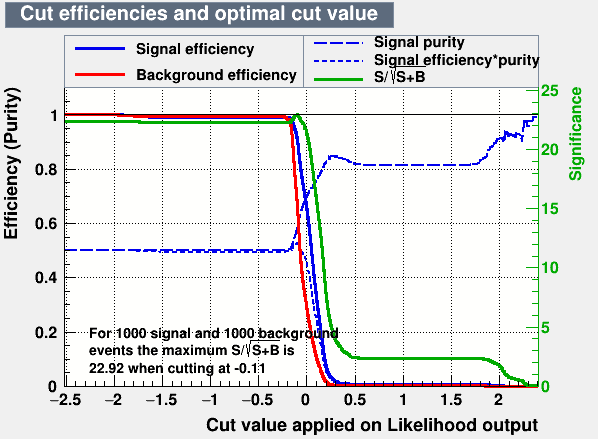}
         \caption{Likelihood Response.}
         \label{fig25b}
     \end{subfigure}
     \hfill
       \caption{The SMLP and Likelihood cut efficiencies as a function of applies cut values respectively.}
        \label{fig18.19}
\end{figure}

Figures~\ref{fig25a} and ~\ref{fig25b} shows the SMLP and Likelihood cut efficiencies as a function of applies cut values respectively. The significance is, 23.16 for BDT, 23.18 for MLP and 22.92 for Likelihood. It has been observed that MLP have higher significance as compared to the BDT and Likelihood method. Kolmogorov-Smirnov test statistics is used in order to determine whether the overtraining occurred. The rule of thumb is used in which we want $\rho_{KS} \ge 0.01$. We have considered two different cases for determining the overtraining, first one is without cuts and second is with cuts.

\begin{figure}[H]
     \centering
     \begin{subfigure}[b]{0.40\textwidth}
         \centering
         \includegraphics[width=\textwidth]{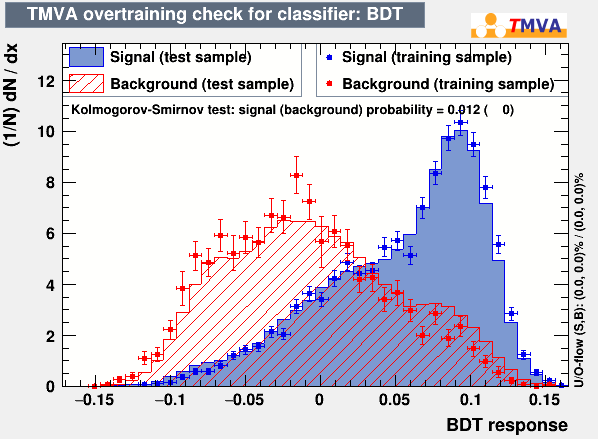}
         \caption{Nomalized BDT response without cuts}
         \label{fig26a}
     \end{subfigure}
     \hfill
     \begin{subfigure}[b]{0.40\textwidth}
         \centering
         \includegraphics[width=\textwidth]{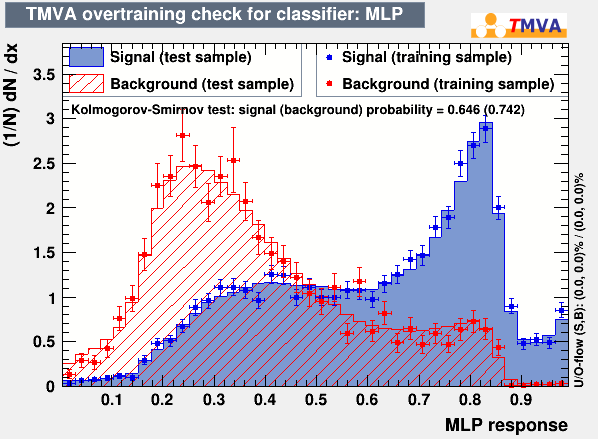}
         \caption{SMLP Response.}
         \label{fig26b}
     \end{subfigure}
     \hfill
     \begin{subfigure}[b]{0.40\textwidth}
         \centering
         \includegraphics[width=\textwidth]{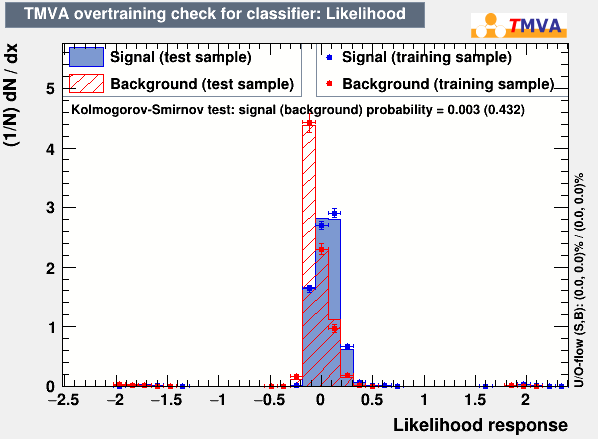}
         \caption{Signal and Background for Likelihood.}
         \label{fig26c}
     \end{subfigure}
        \caption{(a),(b) and (c) represents BDT, MLP and likelihood classifier responses for over-training check.}
        \label{fig26a.26b.26c}
\end{figure}

Figure~\ref{fig26a} shows the normalized BDT response for the samples of test (histograms) and train (dots along with error bars). BDT responses for both signals (blue) and background (red) samples, spreads over the positive and negative BDT response zones respectively. The signals and backgrounds are partially separated from each other and are superimposed. The fact that the KS test statistics are lesser than 0.01 shows that there is overtraining for backgrounds. It can be seen from figure ~\ref{fig26c} that the signal and background histograms almost overlap each other.

Table~\ref{Table5} lists the $K-S$ probability is for signal and background from various methods used for current study. 
\begin{table}[H]
	\centering
	\caption{K-S Probability}
	\hspace*{-2cm}
	\begin{tabular}{|l|l|l|l|l|}
		\hline
		{\bf TMVA} & {\bf Without cuts} &    \\ \hline
		Classifier & Signals  & Background  \\ \hline
		BDT & 0.012 & 0  \\ \hline
		MLP & 0.646 & 0.742 \\ \hline
		LH & 0.003 & 0.432  \\ \hline
	\end{tabular}
	\label{Table5}
\end{table}

\section{Conclusion}
The value of significance of MLP (23.18) is higher than the BDT(23.16) and likelihood (22.92) methods. It concludes that MLP performace is slightly better than BDT and also their ROC curves are also close to each other. Moreover, the signal purity (P) of MLP is splitted earlier and distinguishes the signals ($P \ge 0.5$) from the backgrounds ($P < 0.5$) efficiently. The signal purity response of MLP is much efficient as compared to other two classifiers. 

From the K-S test probability we see that MLP classifier is not overtrained and also after the application of cuts, the overtraining of the signals in MLP classifier decreases, so the MLP outperforms the other two classifiers (BDT and LH) for the classification of signal and backgrounds. It can be concluded that MLP method is to be preferred in our scenario. Three different classifiers are trained to find which one is the most effective in distinguishing signal from backgrounds.

\end{document}